\documentclass[12pt]{article}
\usepackage{epsfig}
\newcommand{\singlespace}{
    \renewcommand{\baselinestretch}{1}\large\normalsize}
\newcommand{\doublespace}{
    \renewcommand{\baselinestretch}{1.6}\large\normalsize}
\newcommand{\bce}{\begin{center}}
\newcommand{\ece}{\end{center}}
\newcommand{\beq}{\begin{equation}}
\newcommand{\eeq}{\end{equation}}
\newcommand{\ba}{\begin{array}}
\newcommand{\ea}{\end{array}}
\newcommand{\bea}{\vspace{0.25cm}\begin{eqnarray}}
\newcommand{\eea}{\end{eqnarray}}
\newcommand{\bfe}{\begin{fmfgraph*}}
\newcommand{\efe}{\end{fmfgraph*}}

\newcommand{\ket}[1]{\vert {#1} \rangle}
\newcommand{\bra}[1]{\langle {#1} \vert}

\newcommand{\dslash}{\partial\!\!\!/}

\newcommand{\qslash}{q\!\!\!/}
\newcommand{\rhoslash}{\rho\!\!\!/}

\newcommand{\elementvon}{{\cal 2}}
\begin{document}
\begin{titlepage}
\vspace{1.0in}
\begin{flushright}
preprint IKDA 98/14
\end{flushright}
\begin{flushright}
July 1998
\end{flushright}
\vspace{1.0in}
\begin{center}
\doublespace
\begin{large}
 {\bf Momentum Dependence of the Pion Cloud for Rho Mesons in Nuclear
  Matter}\\
\end{large}
\vskip 1.0in M. Urban$^1$, M. Buballa$^1$, R. Rapp$^2$ and J. Wambach$^1$\\
{\small
{\it 1) Inst. f. Kernphysik, TU Darmstadt,
Schlo{\ss}gartenstr. 9, 64289 Darmstadt, Germany}\\
{\it 2) Department of Physics, State University of New York,
Stony Brook, NY 11794-3800, U.S.A.}}\\
\end{center}
\vspace{2cm}
\begin{abstract}
We extend hadronic models for $\rho$-meson propagation in cold nuclear matter
via coupling to in-medium pions to include finite three-momentum. Special care
is taken to preserve gauge invariance. Consequences for photoabsorption on the
proton and on nuclei as well as for the dilepton production in relativistic
heavy-ion collisions are discussed.
\end{abstract}
\end{titlepage}
\singlespace
%
\section{Introduction}

\setcounter{page}{2}
The investigation of matter under extreme conditions and the modifications of
hadron properties with density or temperature is one of the main topics in
intermediate and high-energy nuclear physics. Experimentally, the cleanest
information is obtained from electromagnetic probes which penetrate the
medium almost undisturbed. In the vector dominance model
(VDM)~\cite{Sakurai,KLZ} electromagnetic processes are mediated by neutral
vector mesons ($\rho^0$, $\omega$ and $\phi$). Therefore the medium
modifications of vector mesons are of particular interest. For the $\rho$
meson there are both, theoretical and experimental indications that its
properties are changed in nuclear matter:

Using scale invariance arguments, Brown and Rho~\cite{BrownRho} concluded that
the mass of the $\rho$ meson and other hadron masses should drop as a
function of temperature or density. In fact, many authors predicted dropping
vector meson masses, e.g. Saito et al. within the Walecka
model~\cite{SaitoMaruyama} or the Guichon model~\cite{SaitoThomas}.
From a QCD sum rule analysis Hatsuda and Lee~\cite{HatsudaLee} also concluded
that the mass of the $\rho$ meson is shifted downwards considerably at nuclear
matter density. On the other hand, as recently shown by Leupold
et al.~\cite{Leupold}, predictions from QCD sum rules are not unique unless
further assumptions concerning the $\rho$ meson width are made. Klingl et
al.~\cite{Klingl} could satisfy the QCD sum rules with a model which
predicts an increased width of the $\rho$ meson while its mass stays almost
unchanged.

A similar picture emerges from the dilepton data measured by the CERES
collaboration in ultra-relativistic nucleus-nucleus collisions~\cite{CERES}:
the strong enhancement of dilepton production at low invariant masses
($\sim~0.3$ - $0.6$~GeV) was interpreted by Li et al.~\cite{LiKoBrown} as a
signature for a dropping $\rho$-meson mass, supporting the scaling hypothesis
of Brown and Rho~\cite{BrownRho}. Alternatively, however, the CERES dilepton
spectra could also be explained by a strong broadening of the $\rho$ meson
in hot and dense matter~\cite{ChanfrayRappWambach,RappChanfrayWambach}.

There are two classes of ``conventional'' processes which potentially cause
a broadening of the $\rho$ meson: First, as was
demonstrated in the models of Chanfray and Schuck~\cite{ChanfraySchuck}
and of Herrmann et al.~\cite{Herrmann} for cold nuclear matter,
the pion cloud, which gives rise to the vacuum width
of the $\rho$ meson, is modified through interactions with the surrounding
medium. A second important effect is expected from baryonic resonance
formation through direct interactions of the $\rho$ meson with nucleons.
This was considered first by Friman and Pirner~\cite{FrimanPirner} and
investigated extensively by Peters et al.~\cite{Peters}.

Combining both mechanisms, extended to finite temperatures,
Rapp et al.~\cite{RappChanfrayWambach}
obtained a reasonable description of the CERES data. However, the underlying
model for the pionic part~\cite{ChanfraySchuck} was valid only for
$\rho$ mesons at rest ($\vec{q} = 0$). Therefore it was assumed in
ref.~\cite{RappChanfrayWambach}
that this part of the selfenergy does only depend on the invariant mass $M$
of the $\rho$ meson, but not on its 3-momentum $\vec{q}$.
For the resonance contributions the $\vec{q}$-dependence was
taken into account.

The aim of the present paper is to construct a model for the 3-momentum
dependence of the pionic part of the $\rho$-meson selfenergy at zero
temperature. To a large extend we will follow the work of Herrmann et
al.~\cite{Herrmann}. Our article is organized as follows: in
section~\ref{KapRhoVak} we give a short summary of the VDM description of the
$\rho$ meson in vacuum. In section~\ref{KapPiMed} the modification of the
pion propagator in dense matter is discussed. The corresponding corrections
to the $\rho\pi\pi$ and $\rho\rho\pi\pi$ vertices, as required by
Ward-Takahashi identities, will be derived in section~\ref{KapVertKorr}.
Using these vertex functions and the in-medium pion propagator we construct
the $\rho$-meson propagator in matter as detailed in section~\ref{KapRhoMed}.
In section~\ref{KapApp} we discuss two applications of the model. First we
calculate photoabsorption cross sections on nucleons and nuclei and show
how these processes can be used to constrain some of the model parameters,
like the $\pi NN$ form factor. Since real photons have a fixed invariant
mass $M = 0$, the 3-momentum dependence is necessary to obtain nonzero
contributions to the cross section. Finally we will discuss the impact
on dilepton production rates in heavy-ion collisions.
%
%
\section{The $\rho$ Meson in Vacuum}

\label{KapRhoVak}
\setcounter{equation}{0}
The $\rho$ meson has a hadronic width $\Gamma_{\rho}\approx 150$~MeV with
two-pion decay accounting for $\sim$~$100\%$ of it. We restrict ourselves
to the neutral $\rho$ meson, i.e. $\pi^+\pi^-$ decay. The free Lagrangian
involves the isovector pion field $\vec{\phi}$ and the neutral vector field
$\rho^{\mu}$ and reads~\cite{BjorkenDrell}
\beq
{\cal L}_{\pi}+{\cal L}_{\rho} = {1\over 2} \partial_{\mu} \vec{\phi} \cdot
 \partial^{\mu} \vec{\phi} - {1\over 2} m_{\pi}^2 \vec{\phi}
 \cdot \vec{\phi} - {1\over 4} \rho_{\mu\nu} \rho^{\mu\nu}
 + {1\over 2} (m_{\rho}^{(0)})^2 \rho_{\mu} \rho^{\mu}\ ,
\eeq
where $\rho_{\mu\nu} = \partial_{\mu}\rho_{\nu}-\partial_{\nu}\rho_{\mu}$
is the field strength tensor of the $\rho$ field,
and $m_{\rho}^{(0)}$ the bare $\rho$ mass. Minimal substitution,
\beq
\partial_{\mu} \vec{\phi} \longrightarrow
(\partial_{\mu} + ig \rho_{\mu} T_3) \vec{\phi}\ ,
\label{minimalpi}
\eeq
($g$ = $\pi\rho$-coupling constant) leads to the interaction Lagrangian
\beq
{\cal L}_{\pi\rho} = {1\over 2} ig \rho_{\mu} (T_3 \vec{\phi}\cdot
\partial^{\mu}\vec{\phi} + \partial^{\mu}\vec{\phi} \cdot T_3\vec{\phi})
- {1\over 2} g^2 \rho_{\mu} \rho^{\mu} T_3 \vec{\phi} \cdot T_3 \vec{\phi}\ ,
\label{Lpirho}
\eeq
which contains $\rho\pi\pi$ and $\rho\rho\pi\pi$ vertices.

To second order in $g$ the $\rho$-meson selfenergy in vacuum is given by
\bea
-i\Sigma_{\mu\nu}(q)
 &=&g^2\int {d^4k\over{(2\pi)^4}}\,{(2k+q)_{\mu}(2k+q)_{\nu}\over
    {((k+q)^2-m_{\pi}^2+i\varepsilon)(k^2-m_{\pi}^2+i\varepsilon)}}
    \nonumber\\
 & &-2g^2g_{\mu\nu}\int {d^4k\over{(2\pi)^4}}\,{1\over
    {k^2-m_{\pi}^2+i\varepsilon}}\ .
\label{Sigvak}
\eea
The corresponding diagrams are shown in fig.~1.
%
%
\begin{figure}
\begin{center}
\epsfig{file=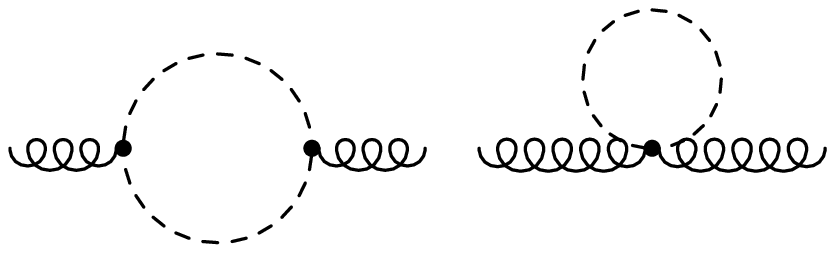,height=3cm,width=8.5cm}
\end{center}
\caption{\small{\it $\rho$-meson selfenergy in vacuum.}}
\end{figure}

In the VDM the $\rho$ meson couples to a conserved current. Consequently,
the selfenergy must vanish for $q^2 = 0$ and has to be 4-dimensionally
transverse:
\beq
q^{\mu}\Sigma_{\mu\nu} = q^{\nu}\Sigma_{\mu\nu} = 0\ .
\label{Transversal}
\eeq
This is in general not fulfilled if the divergent integrals in
eq.~(\ref{Sigvak}) are regularized by a form factor as was done in
ref.~\cite{ChanfraySchuck}. Therefore we will employ the Pauli-Villars
regularization scheme which is known to preserve gauge invariance and
hence transversality. As each integral diverges quadratically, we add
two regulator terms,
\beq
\Sigma_{\mu\nu}(q; m_{\pi})\longrightarrow
\Sigma_{\mu\nu}(q; m_{\pi})+\sum_{i = 1}^{2}c_i\Sigma_{\mu\nu}(q; M_i)\ .
\label{Regul}
\eeq
Here we deviate from ref.~\cite{Herrmann}, where first both integrals are
added, and then the remaining logarithmic divergence is canceled by only one
regulator. Instead we choose
\beq
\begin{array}[b]{ccccccc}
c_1&=&-2\ ,&\quad&M_1&=&\sqrt{m_{\pi}^2+\Lambda_\rho^2}\ ,\\
c_2&=&1\ , &\quad&M_2&=&\sqrt{m_{\pi}^2+2\Lambda_\rho^2}\ ,\\
&&&&&&
\end{array}
\label{Regulwahl}
\eeq
with one free cutoff parameter $\Lambda_\rho$ which is taken as
1 GeV, a typical scale for hadronic models.

Iterating the selfenergy insertions in a Dyson equation by standard
techniques the full $\rho$ propagator of the vacuum model is obtained.
The two remaining parameters, the $\pi\rho$ coupling constant $g$ and the
bare $\rho$ mass $m_{\rho}^{(0)}$ are fitted to the pion electromagnetic
form factor. For $g = 5.9$ and $m_{\rho}^{(0)} = 853$~MeV we obtain the fit
shown in the left panel of fig.~2. With these values the p-wave $\pi\pi$
scattering phase shifts $\delta_1^1$ are also reproduced reasonably well
(right panel of fig.~2). 
%
%
\begin{figure}
\begin{center}
\epsfig{file=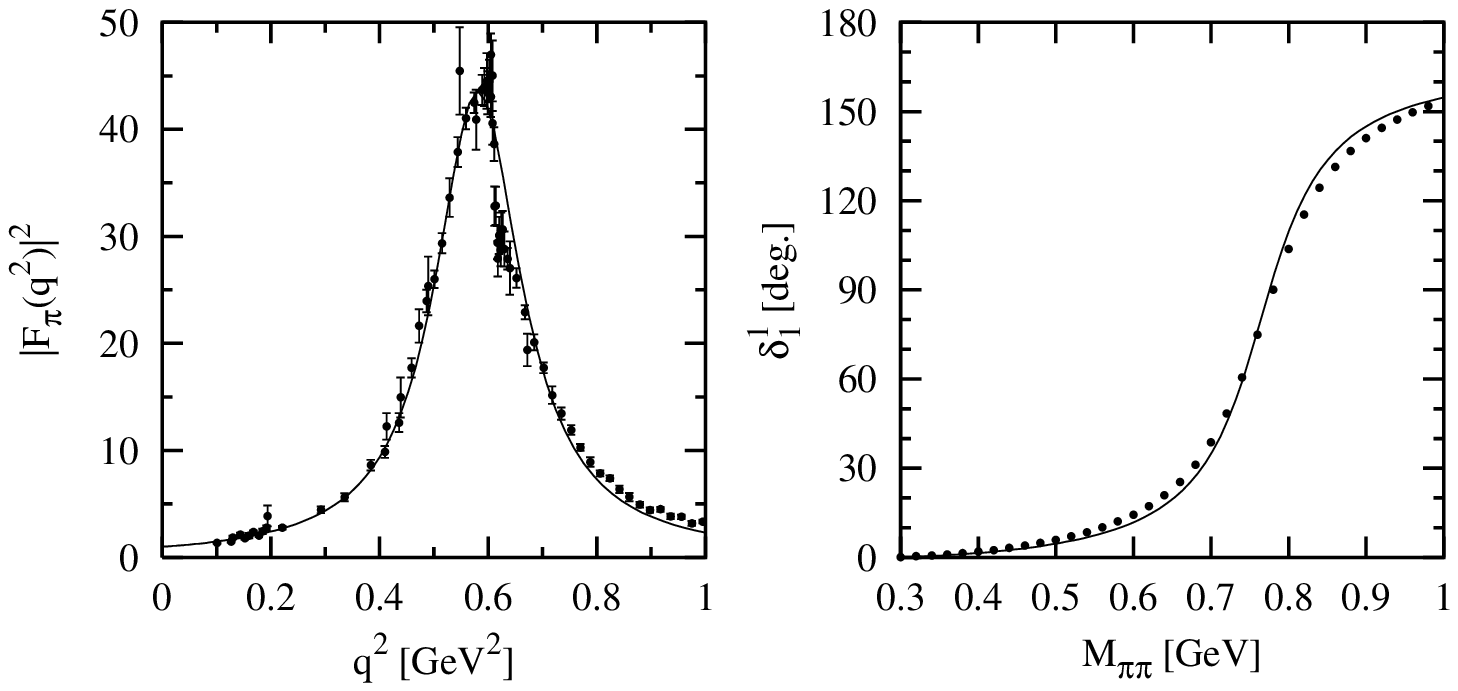,height=7.5cm,width=14.9cm}
\end{center}
\caption{\small {\it Left panel: Electromagnetic form factor of the pion
  $\vert F_{\pi}(q^2)\vert^2$ as a function of $q^2$ in $\mbox{GeV}^2$.
  The data are taken from refs.~\cite{Barkov,Amendolia}.
  Right panel: $\pi\pi$ scattering phase shift $\delta_1^1$ in degrees
  as a function of the invariant mass $M_{\pi\pi}=\sqrt{q^2}$
  in GeV. The data are taken from ref.~\cite{Froggatt}.}}
\end{figure}
%
%
\section{The Pion in Nuclear Matter}

\label{KapPiMed}
\setcounter{equation}{0}
In nuclear matter the interaction of a pion with surrounding nucleons
generates a pion selfenergy. The most important contributions are
particle-hole ($Nh$) and Delta-hole ($\Delta h$) excitations arising from
p-wave pion-nucleon interactions. In the present work they will be treated
in non-relativistic approximation. However, it will turn out to be useful
for the systematic derivation of the coupling to the $\rho$ meson (see next
section) to start from relativistic Lagrangians and then take the
non-relativistic limit.

For the $\pi N$ interaction, we use~\cite{BjorkenDrell,Cubero}:
\bea
{\cal L}_N&=&\bar{\psi}(i\dslash-m_N)\psi\ ,
\label{LNfrei}\\
{\cal L}_{\pi N}&=&{f_N \over {m_{\pi}}}\, \bar{\psi}\gamma^5\gamma^{\mu}
                   \vec{\tau}\psi\cdot \partial_{\mu}\vec{\phi}\ ,
\label{LPiN}
\eea
with a $\pi N$-coupling constant $f_N^2/(4\pi) = 0.08$. The standard
non-relativistic Feynman rules can be derived formally from these
Lagrangians by neglecting antiparticle contributions and expanding
relativistic vertices and spinors to lowest non-vanishing order in
$1/m_N$. However, the relativistic kinematics in the particle
and hole propagators, $\omega_N(\vec{p}) = \sqrt{\vec{p}^{\,2}+m_N^2}$,
will be kept.

For a relativistic treatment of the spin-$3/2$ Delta resonance ($\Delta$),
the formalism of Rarita and Schwinger~\cite{RaritaSchwinger} can be applied.
The free $\Delta$ Lagrangian and the $\pi N\Delta$ interaction Lagrangian
read~\cite{Cubero,RaritaSchwinger}
\bea
{\cal L}_{\Delta}&=&-\bar{\psi}_{\mu}(i\dslash-m_{\Delta})\psi^{\mu}
  +{i\over 3}\bar{\psi}_{\mu}(\gamma^{\mu}\partial_{\nu}
    +\gamma_{\nu}\partial^{\mu})\psi^{\nu}
  -{1\over 3}\bar{\psi}_{\mu}\gamma^{\mu}(i\dslash
    +m_{\Delta})\gamma_{\nu}\psi^{\nu}\ ,
\label{LDeltafrei}\\
{\cal L}_{\pi N\Delta}&=&-{f_{\Delta}\over {m_{\pi}}}
  \bar{\psi}\vec{T}^{\dag}\psi_{\mu}\cdot\partial^{\mu}\vec{\phi}
 \quad +\quad\mbox{h.c.}\ .
\label{LPiNDelta}
\eea
For the $\pi N\Delta$-coupling constant we adopt the Chew-Low value
$f_\Delta = 2 f_N$.
The vertices derived from ${\cal L}_{\pi N\Delta}$ and the Rarita-Schwinger
spinors are expanded in powers of $1/m_{\Delta}$, and again only the
lowest non-vanishing order is retained, leading to standard Feynman rules.
As for the nucleon propagator, we keep relativistic kinematics,
$\omega_{\Delta}(\vec{p}) = \sqrt{\vec{p}^{\,2}+m_{\Delta}^2}$, but neglect
the antiparticle contribution to the $\Delta$ propagator. To account for
the finite $\Delta$ width, $\Gamma_{\Delta}\approx 120$~MeV, we introduce a
constant imaginary part in the denominator of the $\Delta$ propagator.
As will be seen later, inclusion of the momentum and energy dependence of the
delta width would lead to enormous complications in maintaining gauge
invariance of the $\rho$-meson selfenergy.

In this approximation the $Nh$ and $\Delta h$ contributions to the pion
selfenergy are given by
\beq
(\Sigma_{Nh,\,\Delta h})_{ab}(k)=\vec{k}^2\delta_{ab}\Pi_{Nh,\,\Delta h}(k)\ .
\label{SigPiNh}
\eeq
with isospin indices $a$ and $b$, and the dimensionless Lindhard functions
\bea
\Pi_{Nh}(k)
&=&4\Big({f_N\over {m_{\pi}}}\Big)^2 \int_{\vert\vec{p}\vert<p_F}
  {d^3p\over {(2\pi)^3}}\,\Theta(\vert\vec{p}+\vec{k}\vert-p_F)\nonumber \\*
&&\Big({1\over {k^0-\omega_N(\vec{p}+\vec{k})+\omega_N(\vec{p})+i\varepsilon}}
  -{1\over {k^0+\omega_N(\vec{p}+\vec{k})-\omega_N(\vec{p})-i\varepsilon}}
  \Big) \nonumber \\
\label{DefNhLindhard}
\label{PiNh}
\eea
and
\bea
\Pi_{\Delta h}(k)
&=&{16\over 9}\Big({f_{\Delta}\over {m_{\pi}}}\Big)^2
  \int_{\vert\vec{p}\vert<p_F}{d^3p\over {(2\pi)^3}} \nonumber \\*
&&\Big({1\over
    {k^0-\omega_{\Delta}(\vec{p}+\vec{k})+\omega_N(\vec{p})+
      {i\over 2}\Gamma_{\Delta}}}
  -{1\over
    {k^0+\omega_{\Delta}(\vec{p}+\vec{k})-\omega_N(\vec{p})-
      {i\over 2}\Gamma_{\Delta}}}
  \Big)\ . \nonumber \\
\label{DefPiDeltah}
\label{PiDh}
\eea
Because of the relativistic kinematics, only the angular integrations can
be evaluated analytically, whereas the integration over $\vert\vec{p}\vert$
must be performed numerically.

Phenomenology furthermore requires to include repulsive short-range
correlations in the particle-hole bubbles, parametrized in terms of Migdal
parameters $g^{\prime}_{11}$ ($NNNN$ vertex), $g^{\prime}_{12}$ ($NNN\Delta$
vertices) and $g^{\prime}_{22}$ ($NN\Delta\Delta$ vertex) \cite{Migdal}.
These interactions lead to a renormalization of the $\pi NN$ and $\pi N\Delta$
vertices and induce a mixing of the $Nh$ and $\Delta h$ excitations:
\beq
\Pi = {\Pi_{Nh}+\Pi_{\Delta h}
  -(g^{\prime}_{11}-2g^{\prime}_{12}+g^{\prime}_{22})\Pi_{Nh}\Pi_{\Delta h}
  \over {1-g^{\prime}_{11}\Pi_{Nh}-g^{\prime}_{22}\Pi_{\Delta h}
  +(g^{\prime}_{11}g^{\prime}_{22}-g^{\prime 2}_{12})\Pi_{Nh}\Pi_{\Delta h}}}
\ .
\label{PiMigdal}
\eeq
Finally we introduce a monopole form factor at the $\pi NN$ and $\pi N\Delta$
vertex which, in our non-relativistic approximation, depends only on the
three-momentum $\vec{k}$ of the pion,
\footnote{
Note that this form factor is different from
$(\Lambda^2-m_\pi^2)/(\Lambda^2+\vec{k}^2)$, used in
refs.~\cite{ChanfrayRappWambach,RappChanfrayWambach,RappUrbanBuballa},
which was adopted from the Bonn potential~\cite{BonnBibel}. Alternatively
one can choose the form (\ref{PiFormf}) with a redefined coupling constant
$f_{N(\Delta)}(\vec{k}^2=0)$~\cite{BonnBibel}. This choice is more
appropriate in the present context since $\Gamma_{\pi}(\vec{k}=0)$ is
normalized to unity, independent of $\Lambda$.}
\beq
\Gamma_{\pi}(\vec{k}) = {\Lambda^2\over {\Lambda^2+\vec{k}^2}}\ .
\label{PiFormf}
\eeq
The final result for the pion selfenergy thus becomes
\beq
\Sigma^{\prime}_{ab}(k) = \vec{k}^2\delta_{ab}\Pi^{\prime}(k)
               = \vec{k}^2\delta_{ab}\Gamma_{\pi}^2(\vec{k})\Pi(k)\ ,
\label{PimitFF}
\eeq
and the in-medium pion propagator is obtained as
\beq
G_{\pi}(k) = {1\over{k^2-m_{\pi}^2-\Sigma^{\prime}(k)}}
  ={1\over{k^2-m_{\pi}^2-\vec{k}^2\Pi^{\prime}(k)}}\  .
\label{PiPropMed}
\eeq

To render the numerical (in-medium) calculations in sec.~\ref{KapRhoMed}
more tractable most of our
calculations will be performed within the so-called ``3-level model''.
In this approximation, which is equivalent to
neglecting the Fermi motion of the nucleons, the pion propagator is
a mixture of three quasiparticle propagators.
Besides reducing the numerical effort it allows for a more transparent
physical interpretation of the results.

Let us briefly outline the main features of the 3-level model.
We start from eqs.~(\ref{PiNh}) and~(\ref{PiDh}) for the $Nh$ and $\Delta h$
polarization. Assuming the momentum $\vert\vec{p}\vert$ of the hole to be
small compared to the pion momentum $\vert\vec{k}\vert$, one obtains
\beq
\Pi_{Nh}(k)={\alpha_{Nh}(\vec{k})\over{k_0^2-\Omega_{Nh}^2(\vec{k})
  +i\varepsilon}}\ ,\quad
\Pi_{\Delta h}(k)={\alpha_{\Delta h}(\vec{k})\over{k_0^2-
  \Omega_{\Delta h}^2(\vec{k})}}\ ,
\label{PiNhDh3Niveau}
\eeq
where
\bea
\Omega_{Nh}(\vec{k}) = \omega_N(\vec{k})-m_N\ ,&\quad&
\alpha_{Nh}(\vec{k}) = 4 {p_F^3\over{3\pi^2}}
  \Big({f_N\over m_{\pi}}\Big)^2\Omega_{Nh}(\vec{k})\ , \nonumber \\
\Omega_{\Delta h}(\vec{k}) = \omega_{\Delta}(\vec{k})-m_N
  -{i\over 2}\Gamma_{\Delta}\ ,&\quad&
\alpha_{\Delta h}(\vec{k}) = {16\over 9}\,{p_F^3\over{3\pi^2}}
  \Big({f_{\Delta}\over m_{\pi}}\Big)^2\Omega_{\Delta h}(\vec{k})\ .
  \nonumber\\
\eea
Substituting expressions~(\ref{PiNhDh3Niveau}) into eq.~(\ref{PiMigdal}), the
pion selfenergy can be written as
\beq
\Pi(k) = {\alpha_1(\vec{k})\over {k_0^2-\Omega_1^2(\vec{k})}}
        +{\alpha_2(\vec{k})\over {k_0^2-\Omega_2^2(\vec{k})}}\ ,
\eeq
with $\alpha_{1,2}$ and $\Omega^2_{1,2}$ being some combination of
$\alpha_{Nh}$, $\alpha_{\Delta h}$, $\Omega_{Nh}^2$ and $\Omega_{\Delta h}^2$.
Again, this result must be multiplied by $\Gamma_{\pi}^2$:
\beq
\Pi^{\prime}(k) =
  {\alpha^{\prime}_1(\vec{k})\over {k_0^2-\Omega_1^2(\vec{k})}}
 +{\alpha^{\prime}_2(\vec{k})\over {k_0^2-\Omega_2^2(\vec{k})}}\ ,
\label{Piprime3Niveau}
\eeq
where $\alpha^{\prime}_{1,2}(\vec{k}) = \Gamma_{\pi}^2(\vec{k})\,
  \alpha_{1,2}(\vec{k})$. Finally we insert expression~(\ref{Piprime3Niveau})
into eq.~(\ref{PiPropMed}), and obtain a pion propagator of the following
structure:
\beq
G_{\pi}(k) = {S_1(\vec{k})\over {k_0^2-\omega_1^2(\vec{k})}}
            +{S_2(\vec{k})\over {k_0^2-\omega_2^2(\vec{k})}}
            +{S_3(\vec{k})\over {k_0^2-\omega_3^2(\vec{k})}}\ .
\label{PiProp3Niveau}
\eeq
The pion is a mixture of three quasiparticles with
dispersion relations $\omega_1$, $\omega_2$ and $\omega_3$, i.e. the so-called
$(Nh)_L$-branch, the pion-branch and the $(\Delta h)_L$-branch, respectively.
Here the index $L$ means ``longitudinal'', indicating that only
spin-longitudinal excitations (i.e. $\propto \vec{\sigma}\cdot\vec{k}$ or
$\vec{S}\cdot\vec{k}$) can mix with the pion. On the other hand, $\Omega_1$
and $\Omega_2$ can be identified as the dispersion relations for
spin-transverse excitations $(Nh)_T$ and $(\Delta h)_T$ which do not mix
with the pion. Fig.~3 shows these three branches for
$\varrho=\varrho_0=0.16\mbox{ fm}^{-3}$ as well as the dispersion relations
$\Omega_1$, $\omega_{\pi}$ and $\Omega_2$ of non-interacting $Nh$, pion and
$\Delta h$ (quasi-)particles, respectively. The corresponding strength
functions $S_1$, $S_2$ and $S_3$ are displayed in the right panel of fig.~3. 
%
%
\begin{figure}
\begin{center}
\epsfig{file=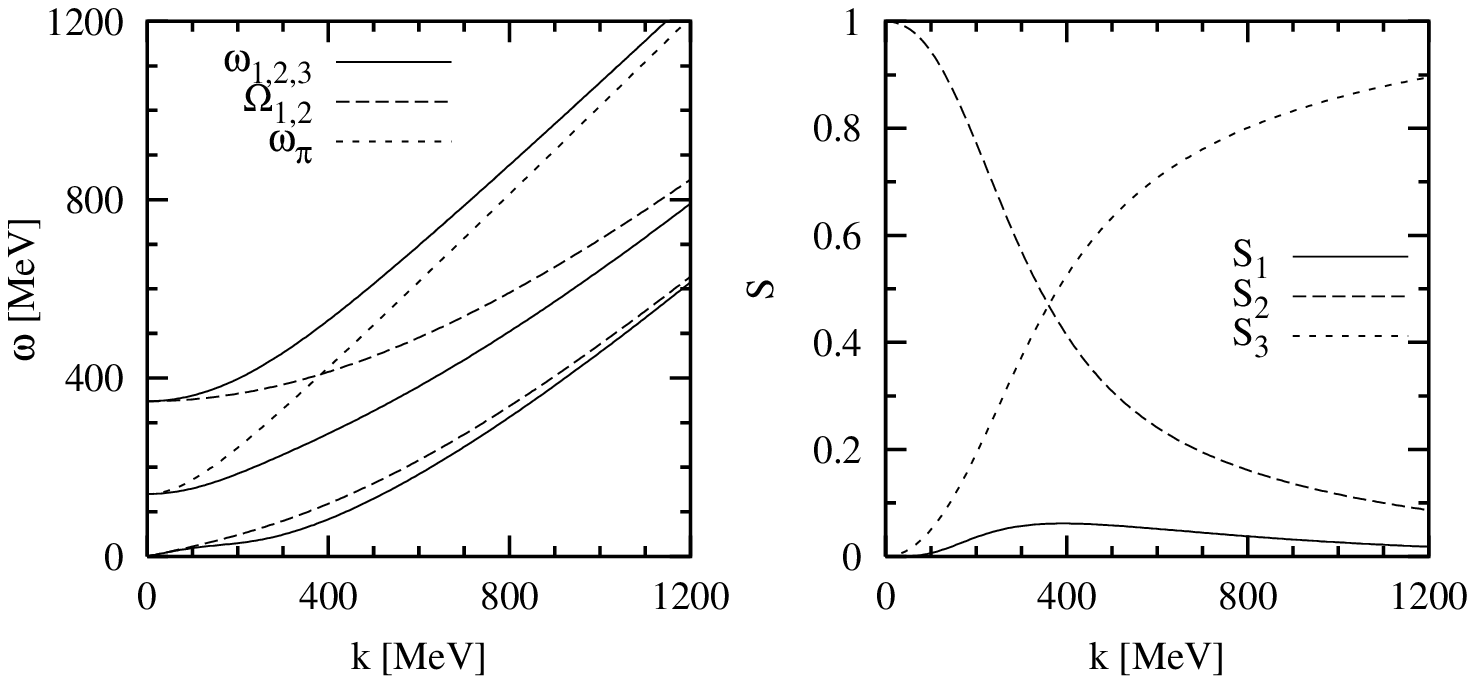,height=7.5cm,width=14.9cm}
\end{center}
\caption{\small {\it Pion properties in the 3-level model calculated at
  nuclear matter density $\varrho=\varrho_0=0.16\mbox{ fm}^{-3}$ using
  $\Lambda = 1200$~MeV, $g'_{11} = 0.8$ and $g'_{12} = g'_{22} = 0.5$
  \cite{ChanfrayRappWambach, RappChanfrayWambach}. Left panel: Dispersion
  relations of the various branches. Solid lines: $\omega_{1,2,3}$ (from
  bottom to top), long-dashed lines: $\Omega_{1,2}$. The short-dashed line
  indicates the dispersion relation of the free pion, $\omega_\pi$.
  Right panel: Strengths of the three branches, $S_1$ (solid line),
  $S_2$ (long-dashed line) and $S_3$ (short-dashed line).}}
\end{figure}
%
%
\section{$\rho\pi\pi$ and $\rho\rho\pi\pi$ Vertex Corrections}

\label{KapVertKorr}
\setcounter{equation}{0}
The transversality of the $\rho$-meson selfenergy can be inferred
from Ward-Takahashi identities which relate the $\rho\pi\pi$ and
$\rho\rho\pi\pi$ vertex functions
to the inverse pion propagator~\cite{Herrmann}:
\bea
q^{\mu}\Gamma_{\mu ab}^{\prime\,(3)}(k,q)
  &=&g\varepsilon_{3ab}\Big(G_{\pi}^{-1}(k+q)-G_{\pi}^{-1}(k)\Big)\ ,
  \label{WardTakahashi3}\\
q^{\mu}\Gamma_{\mu\nu ab}^{\prime\,(4)}(k,k,q)
  &=&ig\Big(\varepsilon_{3ca}\Gamma_{\nu bc}^{\prime\,(3)}(k,-q)-
            \varepsilon_{3bc}\Gamma_{\nu ca}^{\prime\,(3)}(k+q,-q)\Big)\ .
\label{WardTakahashi4}
\eea
In eq.~(\ref{WardTakahashi4}) we have restricted ourselves to the case
of equal in- and outgoing pion momenta ($k_1=k_2$) which is sufficient
for our purposes. The vertex functions $\Gamma_{\mu ab}^{\prime\,(3)}(k,q)$
and $\Gamma_{\mu\nu ab}^{\prime\,(4)}(k_1,k_2,q)$ are illustrated in fig.~4.
%
%
\begin{figure}
\begin{center}
\epsfig{file=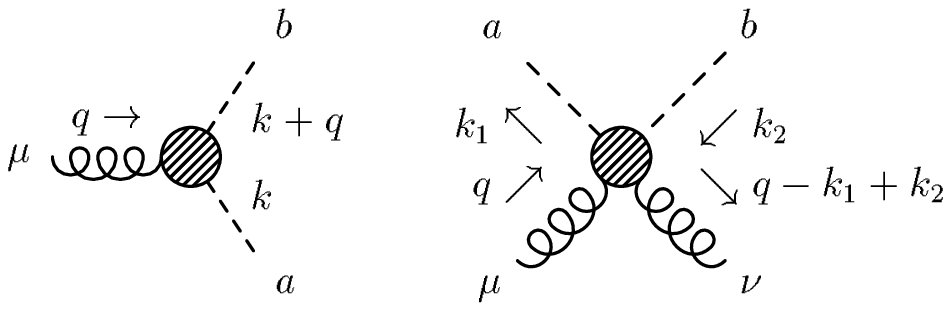,height=3.2cm,width=10cm}
\end{center}
\caption{\small {\it Vertex functions $\Gamma_{\mu ab}^{\prime\,(3)}(k,q)$
  (left) and $\Gamma_{\mu\nu ab}^{\prime\,(4)}(k_1,k_2,q)$ (right).}}
\end{figure}

It is convenient to split the full vertex functions into a bare part
and a vertex correction:
\bea
\Gamma^{\prime\,(3)}_{\mu ab}(k,q) &=& g\varepsilon_{3ab}(2k+q)_{\mu}+
  \tilde{\Gamma}^{\prime\,(3)}_{\mu ab}(k,q)\ , \nonumber \\
\Gamma^{\prime\,(4)}_{\mu\nu ab}(k_1,k_2,q) &=&
  2ig^2(\delta_{ab}-\delta_{3a}\delta_{3b})g_{\mu\nu}+
  \tilde{\Gamma}^{\prime\,(4)}_{\mu\nu ab}(k_1,k_2,q)\ .
\label{vertexcorrection}
\eea
In the vacuum the Ward-Takahshi identities are trivially satisfied. In
nuclear matter, where the pion propagator is modified as discussed in the
previous section, vertex corrections must be taken into account. They can
be constructed by drawing all possible selfenergy insertions for the pion
propagator and then couple the $\rho$ meson to all possible lines and
vertices.

As a first step we will calculate the vertex corrections
$\tilde{\Gamma}^{(3\,Nh)}$ and $\tilde{\Gamma}^{(4 Nh)}$ which correspond
to the pion selfenergy $\Sigma_{Nh}$, i.e. $Nh$ excitations without the
$\pi NN$ form factor. For this purpose we have to specify the interaction
of the $\rho$ meson with nucleons. To ensure gauge invariance, the interaction
Lagrangians are constructed from the Lagrangians ${\cal L}_{N}$,
${\cal L}_{\pi N}$ (eqs.~(\ref{LNfrei}) and (\ref{LPiN})) by minimal
substitutions similar to eq.~(\ref{minimalpi}):
\bea
{\cal L}_{\rho N}&=&-{g\over 2}\bar{\psi}\rhoslash\tau_3\psi\ ,
\label{LrhoN} \\
{\cal L}_{\rho\pi N}&=&ig{f_N \over {m_{\pi}}} \bar{\psi}\gamma^5\rhoslash
   \vec{\tau}\psi\cdot T_3\vec{\phi}\ .
\label{LrhopiN}
\eea
For the resulting vertices we again perform a non-relativistic expansion
keeping only the leading order $(1/m_N)^0$.
The relativistic $\rho NN$ vertex and its non-relativistic expansion read
\beq
\Gamma_{\mu}^{(\rho NN)}
  =-i{g\over 2}\gamma_{\mu}\tau_3
    \longrightarrow\left\{\begin{array}{ll}
      -i{g\over 2}\tau_3+{\cal O}\Big({1\over{m_N^2}}\Big)
        &\quad\mu = 0\ ,\\
      0+{\cal O}\Big({1\over{m_N}}\Big)
        &\quad\mu = 1,2,3\ .
    \end{array}\right.
\label{VertexRhoNNalt}
\eeq
Thus, to order $(1/m_N)^0$, the $\rho$ meson couples only to the nucleon
charge and not to the convection- or magnetization currents. In fact, one
could have added a term to eq.~(\ref{LrhoN}) which describes the tensor
coupling of the $\rho$ meson to the anomalous magnetic moment of the nucleon
and which is gauge invariant by itself. However this term would also not
contribute to order $(1/m_N)^0$.

The relativistic $\rho NN$ vertex fulfills the Ward-Takahashi identity
\beq
q^{\mu}\Gamma^{(\rho NN)}_{\mu}(p,q) = -ig{\tau_3\over 2} \qslash =
  -ig{\tau_3\over 2}\Big(S_F^{-1}(p+q)-S_F^{-1}(p)\Big)\ ,
\eeq
where $p$ is the nucleon momentum  and $q$ the momentum of the
$\rho$ meson. In a consistent non-relativistic approximation an analogous
relation should hold if we replace the relativistic nucleon propagator
$S_F$ by the non-relativistic one, $G_N$. However, starting from the r.h.s.
of eq.~(\ref{VertexRhoNNalt}) the Ward-Takahashi identity is violated at
order $(1/m_N)^1$. To correct for this, we simply add the
missing term to the $0$-component
\beq
\Gamma_{\mu}^{(\rho NN)}
  \longrightarrow\left\{\begin{array}{ll}
    -i{g\over 2}\tau_3{G_N^{-1}(p+q)-G_N^{-1}(p)\over{q^0}}
      +{\cal O}\Big({1\over{m_N}}\Big)&\quad\mu = 0\ ,\\
    0+{\cal O}\Big({1\over{m_N}}\Big) &\quad\mu = 1,2,3\ .\\
  \end{array}\right.
\label{VertexRhoNN}
\eeq
At least for time-like momenta, $q$, this approximation is of the same
accuracy as the non-relativistic expansion. A similar approximation is
necessary for the following expression appearing in some $\rho\rho\pi\pi$
vertex corrections,
\bea
A_{\mu\nu}&=&iG_N(p+q_1-q_2)\Gamma_{\nu}^{(\rho NN)}(p+q_1,-q_2)iG_N(p+q_1)
\Gamma_{\mu}^{(\rho NN)}(p,q_1)iG_N(p) \nonumber\\
&&+iG_N(p+q_1-q_2)\Gamma_{\mu}^{(\rho NN)}(p-q_2,q_1)iG_N(p-q_2)
\Gamma_{\nu}^{(\rho NN)}(p,-q_2)iG_N(p)\ .\nonumber\\
\eea
Using relativistic Feynman rules, i.e. $G_N\rightarrow S_F$ and
$\Gamma_{\mu}^{(\rho NN)}\rightarrow-ig{\tau_3\over{2}}\gamma_{\mu}$,
we find
\beq
q_1^{\mu}q_2^{\nu}A_{\mu\nu}=
-i{{g^2}\over{4}}\Big(S_F(p)+S_F(p+q_1-q_2)-S_F(p+q_1)-S_F(p-q_2)\Big)\ .
\eeq
Within our non-relativistic approximation, the corresponding relation
(with $G_N$ instead of $S_F$ and $\Gamma_{\mu}^{(\rho NN)}$ according to
eq.~(\ref{VertexRhoNN})) is again violated at order
$(1/m_N)^1$. We correct for this again by simply adding the
missing term to the $00$-component:
\beq
A_{\mu\nu}=\left\{\begin{array}{ll}
  -i{{g^2}\over{4}}{1\over{q_1^0q_2^0}}
  \Big(G_N(p)+G_N(p+q_1-q_2)-G_N(p+q_1)-G_N(p-q_2)\Big)&\quad\mu=\nu=0\ ,\\
  0                                             &\quad\mbox{otherwise}\ .\\
  \end{array}\right.
\eeq

From ${\cal L}_{\rho\pi N}$ we deduce the $\rho\pi NN$ vertex as
\beq
  \Gamma^{(\rho\pi NN)\,\mu}_a
  =ig{f_N\over {m_{\pi}}}\varepsilon_{3ba}\gamma^5\gamma^{\mu}\tau_b
  \longrightarrow\left\{\begin{array}{ll}
  0+{\cal O}\Big({1\over{m_N}}\Big) &\quad\mu = 0\ ,\\ 
  -ig{f_N\over {m_{\pi}}}\varepsilon_{3ba}\sigma^{\mu}\tau_b
    +{\cal O}\Big({1\over{m_N^2}}\Big)&\quad\mu = 1,2,3\ .
   \end{array}\right.
\label{VertexRhoPiNN}
\eeq

We are now in place to calculate the vertex corrections corresponding to the 
$Nh$ selfenergy of the pion. Two of the four possible $\rho\pi\pi$ vertex 
correction diagrams are shown in fig.~5,
the remaining ones are obtained by interchanging the pion lines. 
For the $\rho\rho\pi\pi$ vertex corrections, there
exist sixteen possibilities to attach two $\rho$-meson lines to the $Nh$
bubble. Eight of these diagrams are shown in fig.~6 
and the others can be generated by interchanging the pion lines. 
%
%
\begin{figure}
\begin{center}
\epsfig{file=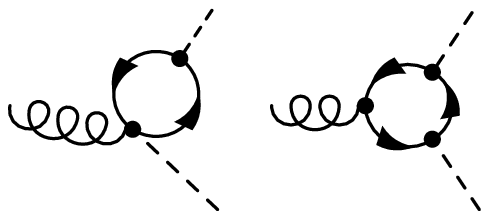,height=2.2cm,width=5.1cm}
\end{center}
\caption{\small {\it $\rho\pi\pi$ vertex correction diagrams due to 
$Nh$-loops.}}
\end{figure}
%
%
\begin{figure}
\begin{center}
\epsfig{file=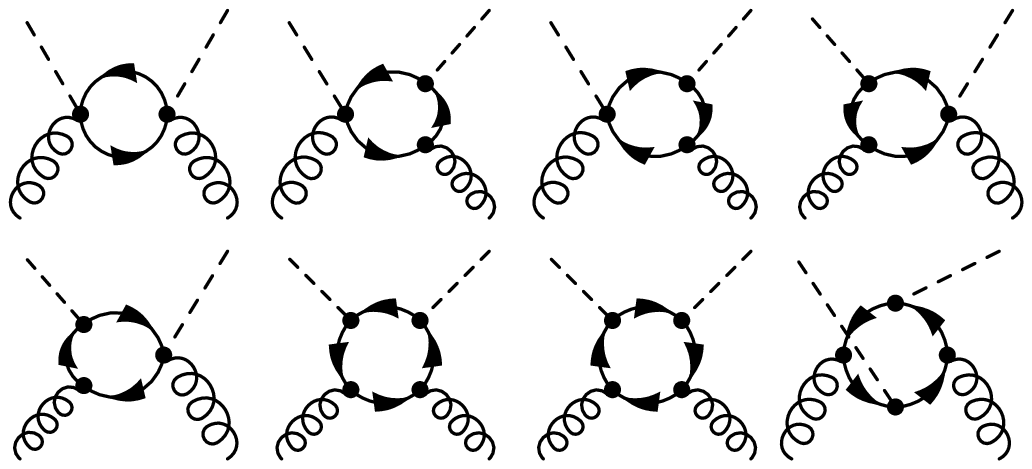,height=4.7cm,width=10.5cm}
\end{center}
\caption{\small {\it $\rho\rho\pi\pi$ vertex correction diagrams due to 
$Nh$-loops.}}
\end{figure}

Using eqs.~(\ref{VertexRhoNN}) and~(\ref{VertexRhoPiNN}) and the
definition~(\ref{DefNhLindhard}), we obtain the following correction for
the $\rho\pi\pi$ vertex function 
\bea
\tilde{\Gamma}^{(3\;Nh)}_{0ab}(k,q)&=&
-g\varepsilon_{3ab}{\vec{k}\cdot(\vec{k}+\vec{q})\over{q^0}}
  \Big(\Pi_{Nh}(k+q)-\Pi_{Nh}(k)\Big)\ , 
\nonumber\\
\tilde{\Gamma}^{(3\;Nh)}_{iab}(k,q)&=&
g\varepsilon_{3ab}\Big((k+q)_i \Pi_{Nh}(k+q)
  +k_i\Pi_{Nh}(k)\Big)\ , \nonumber\\
\label{Gamma3Nh}
\eea
and the $\rho\rho\pi\pi$ vertex correction reads as
\bea
\tilde{\Gamma}_{00ab}^{(4\;Nh)}(k,k,q)&=&
  -ig^2(\delta_{ab}-\delta_{3a}\delta_{3b})\nonumber
  {\vec{k}^2\over {(q^0)^2}}
  \Big(\Pi_{Nh}(k+q)+\Pi_{Nh}(k-q)-2\Pi_{Nh}(k)\Big)\ ,
\nonumber\\
\tilde{\Gamma}_{j0ab}^{(4\;Nh)}(k,k,q)&=&
  -ig^2(\delta_{ab}-\delta_{3a}\delta_{3b}){k_{j}\over q^0}
  \Big(\Pi_{Nh}(k-q)-\Pi_{Nh}(k+q)\Big)\;=\;
\tilde{\Gamma}_{0jab}^{(4\;Nh)}(k,k,q)\ ,
\nonumber\\
\tilde{\Gamma}_{ijab}^{(4\;Nh)}(k,k,q)&=&
  -ig^2(\delta_{ab}-\delta_{3a}\delta_{3b})\delta_{ij}
  \Big(\Pi_{Nh}(k-q)+\Pi_{Nh}(k+q)\Big)\ .\nonumber\\
\label{Gamma4Nh}
\eea

Since the result can be expressed through the pion selfenergy $\Pi_{Nh}$, it 
is quite obvious that the Ward-Takahashi identities~(\ref{WardTakahashi3})
and~(\ref{WardTakahashi4}) are fulfilled for a pion propagator dressed by
$Nh$-loops.

The vertex corrections which correspond to the $\Delta h$ part of the
pion selfenergy can be obtained in the same way. Here we start from the
interaction Lagrangians
\bea
{\cal L}_{\rho\Delta} &=&
  g\bar{\psi}_{\mu}\rhoslash T_3^{({3\over 2})}\psi^{\mu}
  -{g\over 3}\bar{\psi}_{\mu}(\gamma^{\mu}\rho_{\nu}+\gamma_{\nu}\rho^{\mu})
  T_3^{({3\over 2})}\psi^{\nu}
  +{g\over 3}\bar{\psi}_{\mu}\gamma^{\mu}\rhoslash T_3^{({3\over 2})}
  \gamma_{\nu}\psi^{\nu}\ ,\\
{\cal L}_{\rho\pi N\Delta} &=&
  -ig{f_{\Delta}\over{m_{\pi}}}\bar{\psi}\vec{T}^{\dag}\psi_{\mu}\rho^{\mu}
  \cdot T_3\vec{\phi}\quad +\quad\mbox{h.c.}\ , 
\eea
which one gets from gauging ${\cal L}_{\Delta}$ and 
${\cal L}_{\pi N\Delta}$ (eqs.~(\ref{LDeltafrei}) and (\ref{LPiNDelta})).
As stated in the previous section we include a finite width for the 
$\Delta$.
Here one has to be careful not to violate the Ward-Takahashi identity
which relates the $\rho\Delta\Delta$ vertex to the inverse $\Delta$ propagator.
This is the reason why we have restricted ourselves to a constant $\Delta$ 
width which drops out if we take the difference between two inverse 
propagators.

Repeating the entire calculation for the $\Delta h$ diagrams, the result is
completely analogous to 
eqs.~(\ref{Gamma3Nh}) and~(\ref{Gamma4Nh}) with the index $Nh$
replaced by $\Delta h$. For the $\rho\pi\pi$ vertex this has already been
found by Chanfray and Schuck (eq.~(3.7) in ref.~\cite{ChanfraySchuck}).

The next problem is to include the short-range Migdal interaction. Since the
$NNNN$, $NNN\Delta$ and $NN\Delta\Delta$ vertices are momentum independent,
the $\rho$ meson does not couple to these vertices. Nevertheless many
additional diagrams must be calculated. An example for this class of diagrams
is shown in fig.~7. It turns out that relations analogous to 
eqs.~(\ref{Gamma3Nh}) and~(\ref{Gamma4Nh}) are valid for each pion
selfenergy contribution and the corresponding vertex corrections. Hence the
index $Nh$ in the eqs.~(\ref{Gamma3Nh}) and~(\ref{Gamma4Nh}) may be
omitted.
%
%
\begin{figure}
\begin{center}
\epsfig{file=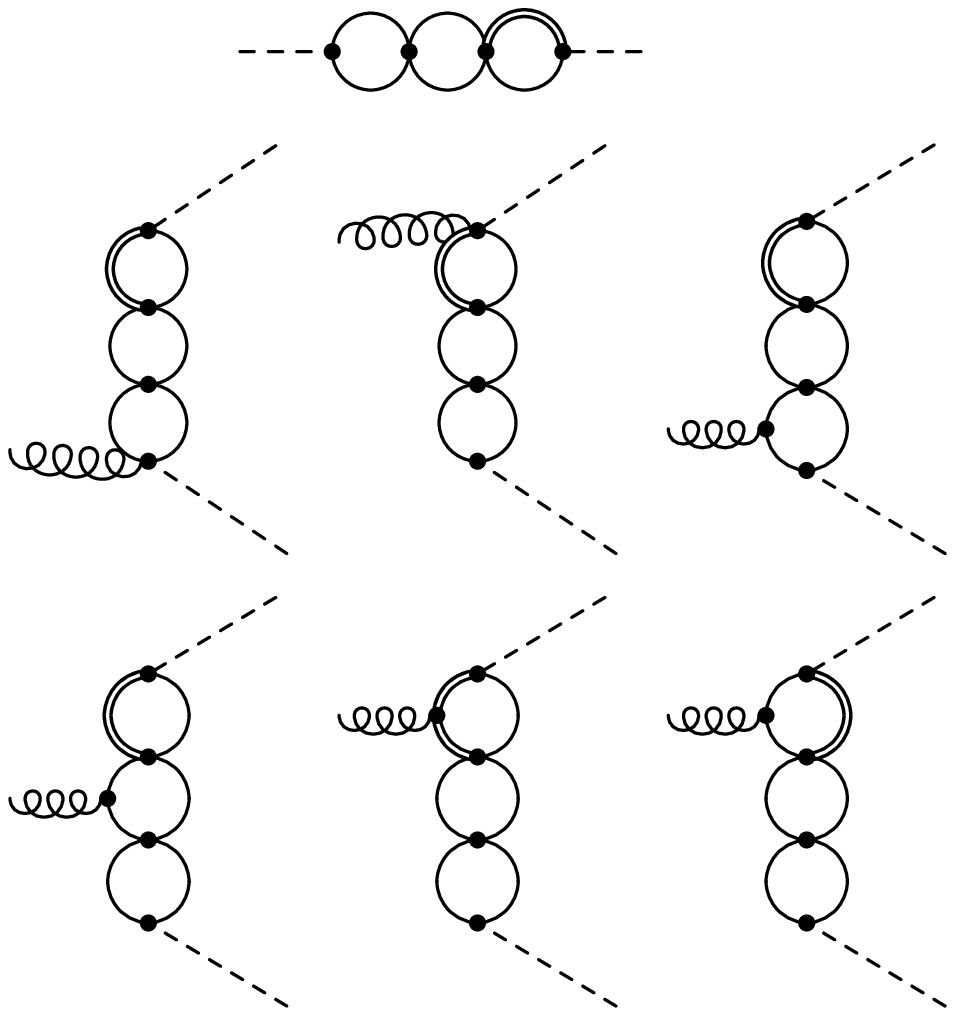,height=10.3cm,width=9.5cm}
\end{center}
\caption{\small {\it One contribution to the pion selfenergy and
  the corresponding $\rho\pi\pi$ vertex corrections.}}
\end{figure}

So far the $\pi NN$ and $\pi N\Delta$ form factor $\Gamma_{\pi}$ has not been 
included. Naively one would expect
that this form factor multiplies the $\rho\pi\pi$ vertex corrections by
$\Gamma_{\pi}(\vec{k})\Gamma_{\pi}(\vec{k}+\vec{q})$ and the $\rho\rho\pi\pi$
vertex corrections by $\Gamma_{\pi}(\vec{k}_1)\Gamma_{\pi}(\vec{k}_2)$.
However, this violates the Ward-Takahashi identities. The reason is that the
$\pi NN$ ($\pi N\Delta$) form factor implies that the $\rho\pi NN$
($\rho\pi N\Delta$) vertex must be modified. We shall do this in a systematic
way as proposed in refs.~\cite{Herrmann,Mathiot}.

Formally the form factor defined in eq.~(\ref{PiFormf}) can be generated by 
the ``propagator'' $1/(-\vec{k}^2-\Lambda^2)$ of a heavy particle
carrying the quantum numbers of a pion (therefore called ``heavy pion''),
which is inserted between the physical pion propagator and the $\pi NN$
($\pi N\Delta$) vertex (see fig.~8). Note that we have omitted the 
$k_0^2$ in the heavy-pion propagator, because $\Gamma_{\pi}$ depends only on
the three momentum $\vec{k}$. In order to get the correct normalization of the
form factor, the ``vertex'' where the pion is converted into the heavy pion 
must be assigned a factor $i\Lambda^2$.
%
%
\begin{figure}
\begin{center}
\epsfig{file=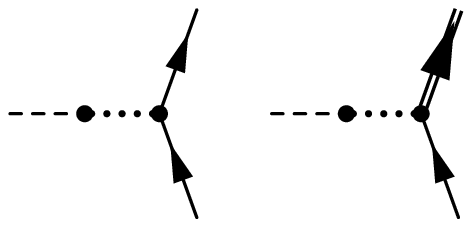,height=2.3cm,width=5cm}
\end{center}
\caption{\small {\it Insertion of the ``heavy pion propagator'' (dotted line) 
to generate the $\pi NN$ and $\pi N\Delta$ monopole form factor.}}
\end{figure}

In this formalism the pion selfenergy $\Sigma^\prime$ corresponds to the 
diagram shown in the left part of fig.~9. Here $\Sigma$ again denotes the
selfenergy without any form factor. Now, according to our general
prescription, the $\rho\pi\pi$-vertex correction can be constructed by
coupling the $\rho$ meson to this diagram in all possible ways. This leads
to the three diagrams which are shown in fig.~9. The first one corresponds to 
the naive expectation that the vertex correction with formfactor is just
the vertex correction calculated without formfactor multiplied by 
$\Gamma_{\pi}(\vec{k})\Gamma_{\pi}(\vec{k}+\vec{q})$. However, there are
two more diagrams where the $\rho$ meson directly couples to the ``heavy
pion'' and which are important for preserving gauge invariance. 
%
%
\begin{figure}
\begin{center}
\epsfig{file=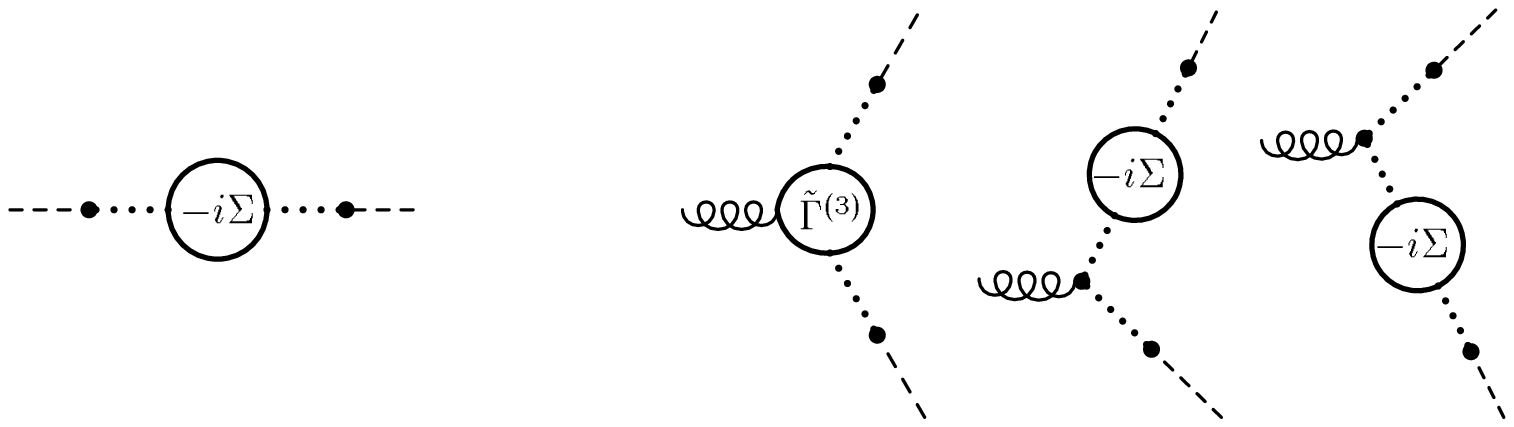,height=5.4cm,width=15.6cm}
\end{center}
\caption{\small {\it Pion selfenergy (left) and $\rho\pi\pi$ vertex 
  corrections with $\pi NN$ ($\pi N\Delta$) form factor generated by the
  ``heavy pion''.}}
\end{figure}

The vertex of the $\rho$ meson coupling to the heavy pion is almost identical
to the $\rho\pi\pi$ vertex. However, since we have omitted the $k_0^2$ in the
heavy pion propagator, we must also set the $\mu=0$ component of the vertex
equal to zero in order to satisfy the Ward-Takahashi identity. Thus the two 
extra diagrams of fig.~9 contribute to the spatial components of the vertex
correction only. The final result is
\beq
\tilde{\Gamma}^{\prime\,(3)}_{\mu ab}(k,q)
=\bigg(\tilde{\Gamma}^{(3)}_{\mu ab}(k,q)
    -g\varepsilon_{3ab}(2k+q)_i
      \Big({\Sigma(k+q)\over{\Lambda^2+(\vec{k}+\vec{q})^2}}+
           {\Sigma(k)\over{\Lambda^2+\vec{k}^2}}\Big)\bigg)
\Gamma_{\pi}(\vec{k})\Gamma_{\pi}(\vec{k}+\vec{q})\ ,
\label{Gamma3FF}
\eeq
where we introduced the following short-hand notation: for $\mu=0$
the term with the index $i$ should be dropped, otherwise $i=\mu$.

In the same way we can construct the $\rho\rho\pi\pi$-vertex correction with
a form factor. Again starting from the pion selfenergy as shown in fig.~9
and coupling two $\rho$ mesons to this diagram we 
find 13 diagrams to evaluate. The result is
\bea
\tilde{\Gamma}^{\prime \,(4)}_{\mu\nu ab}(k,k,q)
&=\Bigg\{&
  \tilde{\Gamma}^{(4)}_{\mu\nu ab}(k,k,q) \nonumber\\
&&-ig\varepsilon_{3ca}\Big[{(2k-q)_i\over{\Lambda^2+(\vec{k}-\vec{q})^2}}
  \tilde{\Gamma}^{(3)}_{\nu bc}(k,-q) 
  + {(2k+q)_j\over{\Lambda^2+(\vec{k}+\vec{q})^2}}
  \tilde{\Gamma}^{(3)}_{\mu cb}(-k-q,q) \nonumber \Big]\\
&&-ig\varepsilon_{3bc} \Big[{(2k+q)_i\over{\Lambda^2+(\vec{k}+\vec{q})^2}}
  \tilde{\Gamma}^{(3)}_{\nu ac}(-k,-q) 
  + {(2k-q)_j\over{\Lambda^2+(\vec{k}-\vec{q})^2}}
  \tilde{\Gamma}^{(3)}_{\mu ca}(k-q,q) \Big] \nonumber\\
&&-ig^2(\delta_{ab}-\delta_{3a}\delta_{3b})
   \Big[{(2k-q)_i(2k-q)_j \over{\Lambda^2+(\vec{k}-\vec{q})^2}}
   \Big({\Sigma(k-q) \over{\Lambda^2+(\vec{k}-\vec{q})^2}}  
       +2 {\Sigma(k) \over{\Lambda^2+\vec{k}^2}} \Big) \nonumber\\
&&  \hspace{32mm}+{(2k+q)_i(2k+q)_j \over{\Lambda^2+(\vec{k}+\vec{q})^2}}
   \Big({\Sigma(k+q) \over{\Lambda^2+(\vec{k}+\vec{q})^2}}  
       +2 {\Sigma(k) \over{\Lambda^2+\vec{k}^2}} \Big) \nonumber\\
&&  \hspace{32mm}-4 \delta_{ij} {\Sigma(k)\over{\Lambda^2+\vec{k}^2}} \Big]
   \quad\Bigg\}\Gamma_{\pi}^2(\vec{k})\ . \nonumber\\
\eea
The meaning of the index $i$ is the same as in eq.~(\ref{Gamma3FF}).
Similarly $j=\nu$ for $\nu=1,2,3$,  whereas the corresponding
terms vanish for $\nu=0$.  
%
%
\section{The $\rho$ Meson in Nuclear Matter}

\label{KapRhoMed}
\setcounter{equation}{0}
We are now ready to address the central task of our article which is the 
description of the $\rho$ meson in dense matter. For this we generalize 
the vacuum selfenergy of the $\rho$ meson as shown in fig.~1 
by replacing the vacuum pion propagators and vertices by the in-medium
propagators and vertex functions as calculated in the previous two sections:
\bea
-i\Sigma_{\mu\nu}(q) &=&
  {1\over 2}\int{d^4 k\over {(2\pi)^4}}
  iG_{\pi}(k)\Gamma_{\mu ab}^{\prime\,(3)}(k,q)iG_{\pi}(k+q)
  \Gamma_{\nu ba}^{\prime\,(3)}(k+q,-q) \nonumber\\*
&&+{1\over 2}\int{d^4 k\over {(2\pi)^4}}iG_{\pi}(k)
  \Gamma_{\mu\nu aa}^{\prime\,(4)}(k,k,q)\ .
\label{SigRhoAllgemein}
\eea
The corresponding diagrams are shown in fig.~10. 
It is easy to see that $\Sigma_{\mu\nu}$ fulfills the transversality condition,
eq.~(\ref{Transversal}), \cite{Herrmann}: using the Ward-Takahashi identities,
eqs.~(\ref{WardTakahashi3}) and (\ref{WardTakahashi4}) one gets 
\bea
-iq^{\mu}\Sigma_{\mu\nu}(q)
&=&{g\over 2}\varepsilon_{3ab}\int{d^4 k\over {(2\pi)^4}}
  \Big(G_{\pi}(k+q)\Gamma_{\nu ba}^{\prime\,(3)}(k+q,-q)
  -G_{\pi}(k)\Gamma_{\nu ba}^{\prime\,(3)}(k,-q)\Big)\ .
\eea
Since a Pauli-Villars regularization is used we are allowed to substitute 
$k\longrightarrow k+q$ in the second term and the entire expression 
vanishes. (Obviously this would not be true if the integral was regularized
by a form factor.)
%
%
\begin{figure}
\begin{center}
\epsfig{file=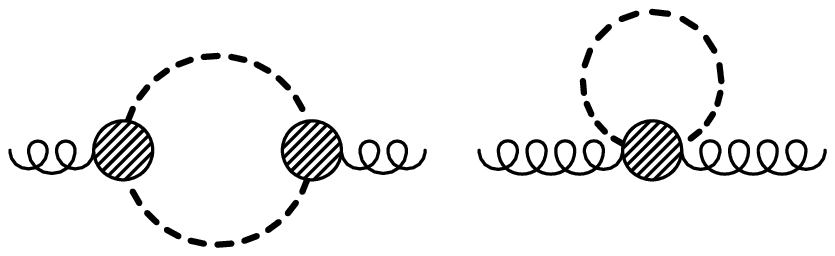,height=3cm,width=8.5cm}
\end{center}
\caption{\small{\it $\rho$-meson selfenergy in nuclear matter.}}
\end{figure}

The transversality of $\Sigma_{\mu\nu}$ implies its general
structure to be~\cite{GaleKapusta} 
\beq
\Sigma_{\mu\nu}(q) = \Sigma_T(q)(P_T)_{\mu\nu}+\Sigma_L(q)(P_L)_{\mu\nu}
\ ,
\eeq
with the transverse and longitudinal projection operators
\beq
P_T^{\mu\nu} =    \left\{\begin{array}{ll} 0
        &\qquad\mu = 0 \quad {\rm or}\quad\nu=0 \\
         \delta^{\mu\nu} - \frac{q^{\mu} q^{\nu}}{\vec{q}^{\,2}}
        &\qquad\mu,\nu~\elementvon~\{ 1,2,3\}
    \end{array}\right.
\quad , \qquad
P_L^{\mu\nu} = \frac{q^\mu q^\nu}{q^2} - g^{\mu\nu} - P_T^{\mu\nu} \;.
\eeq
They are both four-dimensionally transverse, but $P_T$ is
three-dimensionally transverse, where\-as $P_L$ is three-dimensionally 
longitudinal. Since in matter Lorentz invariance is not manifest because there
is a preferred frame of reference in which the matter is at rest,
the two functions $\Sigma_T$ and $\Sigma_L$ depend on $q^0$ and
$\vert\vec{q}\vert$ separately. 

Similarly, the in-medium $\rho$ propagator can be expressed via the
projection operators $P_T$ and $P_L$ as
\bea
G_{\rho}^{\mu\nu}(q)
 &=&G_{\rho\,T}(q) P_T^{\mu\nu}+G_{\rho\,L}(q) P_L ^{\mu\nu}
  +{q^{\mu}q^{\nu}\over {(m_{\rho}^{(0)})^2q^2}}\nonumber\\*
 &=&{P_T^{\mu\nu}\over{q^2-(m_{\rho}^{(0)})^2-\Sigma_T(q)}}
  +{P_L^{\mu\nu}\over{q^2-(m_{\rho}^{(0)})^2-\Sigma_L(q)}}
  +{q^{\mu}q^{\nu}\over {(m_{\rho}^{(0)})^2q^2}}\ .
\label{Grhomunu}
\eea

We can choose the frame of reference such that $\vec{q}$ points in the 
direction of the $z$-axis. In this case $\Sigma_T=\Sigma_{11}$ and
$\Sigma_L=(q^2/q_0^2)\Sigma_{33}$, i.e. it is sufficient to 
work with the spatial components of $\Sigma_{\mu\nu}$.
Therefore we have to insert the spatial components of the vertex functions
calculated in sec.~\ref{KapVertKorr} into eq.~(\ref{SigRhoAllgemein}).

First the $k_0$-integration is performed.  As in eq.~(\ref{vertexcorrection})
we split the vertex functions into a bare part and a vertex correction.
The spatial components of the bare vertices do not depend on $k_0$. Hence, for
these terms the integrands depend on $k_0$ through the pion propagators
$G_\pi$ only, i.e. we have to evaluate the expressions
\beq
I_1={i\over\pi}\int dk^0\,G_{\pi}(k)G_{\pi}(k+q)
\eeq
and
\beq
I_5={i\over\pi}\int dk^0\,G_{\pi}(k)\ .
\eeq
The vertex corrections depend on $k_0$ through the pion selfenergies 
$\Pi^\prime$. This leads to additional integrals which contain pion
propagators and selfenergies, for instance
\beq
I_6={i\over\pi}\int dk^0\,\Pi^{\prime}(k)G_{\pi}(k+q)\ ,
\eeq
which arises from a vertex correction to the $\rho\rho\pi\pi$-vertex
(see fig.~11). Altogether we have to evaluate seven $k_0$-integrals $I_m$,
$m=1,...,7$ which are listed in appendix~\ref{AppGen}. The spatial components
of $\Sigma_{\mu\nu}$ can then be written as
\beq
\Sigma_{ij}(q) =
  {g^2\over{2}}\;\sum_{m=1}^7\;
  \int {d^3k\over{(2\pi)^3}} \;f_{ij}^{(m)}(\vec{k},\vec{q})
  \;I_m(\vec{k},\vec{q},q_0) 
\label{SigRhoMedijgen}
\eeq
with purely real functions $f_{ij}^{(m)}(\vec{k},\vec{q})$. The explicit
expressions are also given in appendix~\ref{AppGen}.
%
%
\begin{figure}
\begin{center}
\epsfig{file=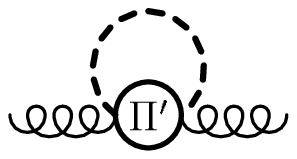,height=2.2cm,width=3cm}
\end{center}
\caption{\small{\it Diagram representing the
  $I_6$ terms in eq.~(\ref{SigRhoMedijgen}).}}
\end{figure}

As mentioned earlier, the computation of $\Sigma_{ij}$ becomes quite involved
if we use the exact pion selfenergies and propagators as defined in 
eqs.~(\ref{SigPiNh}) - (\ref{PiPropMed}). Since these functions are obtained  
numerically, also the $k_0$-integrations for the functions $I_1$ to
$I_7$ have to be performed numerically. Finally they have to be integrated 
over $|\vec{k}|$ and one angle.

Therefore we begin with the 3-level model introduced in the last paragraph
of sec.~\ref{KapPiMed} and will come back to the exact case later. In the
3-level model the selfenergies and propagators are given analytically
(see eqs.~(\ref{PiNhDh3Niveau}) - (\ref{PiProp3Niveau})) and also allow
for an analytical evaluation of the integrals $I_1$ to $I_7$. The results
are listed in appendix~\ref{App3Niveau}.
The remaining integrations in eq.~(\ref{SigRhoMedijgen}), i.e. the integration
over the angle $\vartheta$ between $\vec{k}$ and $\vec{q}$ and the integration
over $\vert\vec{k}\vert$, are performed numerically. 

Results for the $\rho$-meson selfenergy as a function of the invariant 
mass $M=\sqrt{q^2}$ are shown in fig.~12 for different
densities $\varrho$ and momenta $\vert\vec{q}\vert$. The real parts of 
$\Sigma_L$ and $\Sigma_T$ are displayed in the upper three panels. 
As discussed in ref.~\cite{Herrmann}, in matter, they do no longer vanish at 
$M = 0$ because of screening effects. The corresponding imaginary parts
are shown in the lower three panels of fig.~12. Whereas
in vacuum $\mbox{Im}\;\Sigma_{\mu\nu}$ vanishes below the two-pion threshold,
$M = 2m_\pi$, in matter there are new decay channels, such as $Nh$
and $\Delta h$ states, which allow for finite imaginary parts at lower 
invariant masses and even for $M=0$. On the other hand, the imaginary part
of the $\rho$-meson selfenergy should vanish for $q^0\rightarrow 0$,
independently of $\vec{q}$. In our model this is slightly violated because
of the constant width of the $\Delta$. An energy- and momentum independent
$\Delta$ width means that there is a finite probability to find a $\Delta$
with the mass of a nucleon or even below. This leads to 
(unphysical) decay channels of the $\rho$ meson with zero energy which are 
responsible for the non-vanishing selfenergy at $q^0=0$. However, the 
effect is very small. For instance, for $\varrho = 2\varrho_0$ and
$|\vec{q}|=q^0=0$ we find $\mbox{Im}\;\Sigma = 10^{-3} {\rm GeV}^2$ which is
not visible in fig.~12. 
%
%
\begin{figure}
\begin{center}
\epsfig{file=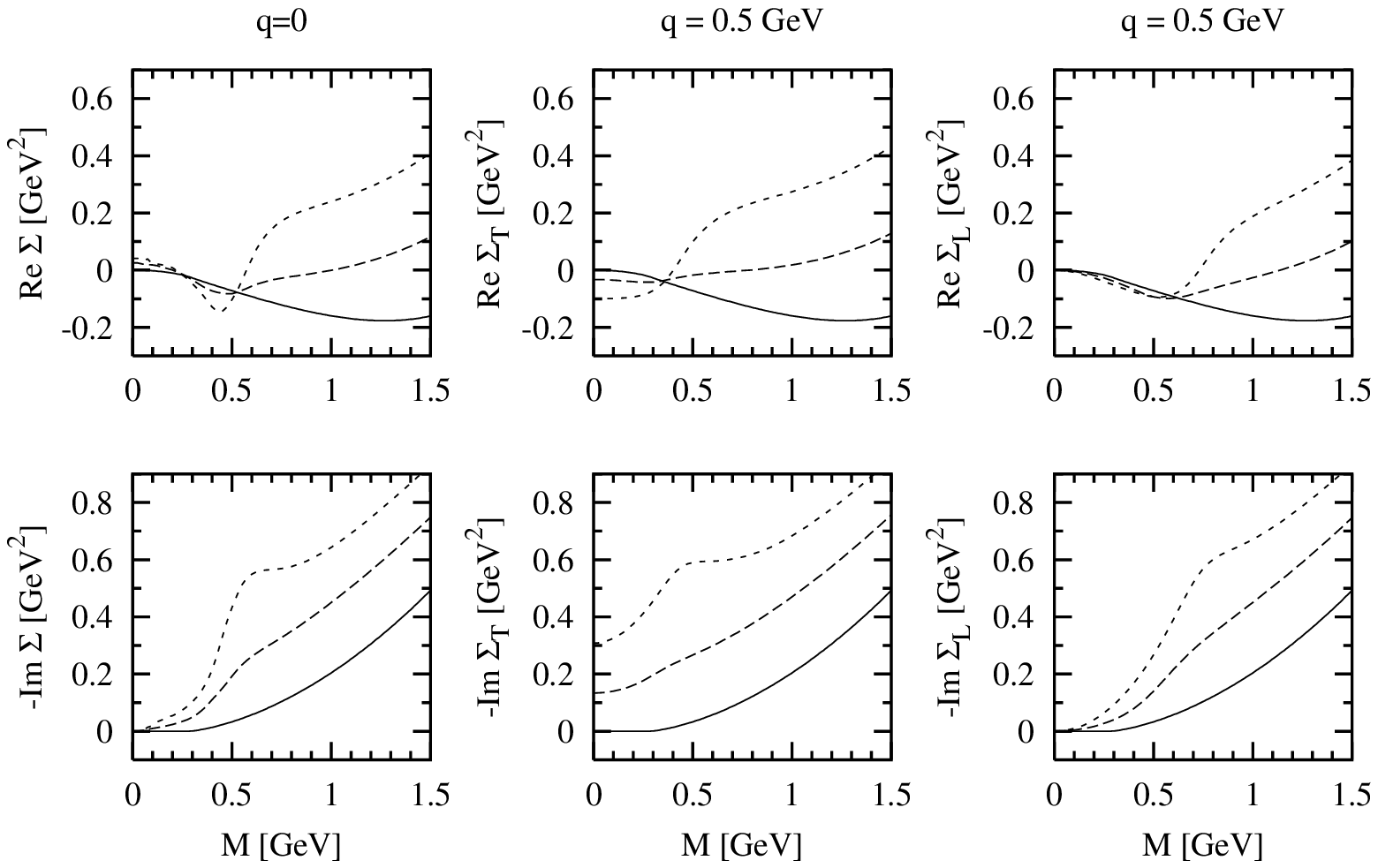,height=11cm,width=15.8cm}
\end{center}
\caption{\small {\it From left to right: $\Sigma = \Sigma_T = \Sigma_L$ for
  $\vert\vec{q}\vert = 0$, $\Sigma_T$ and $\Sigma_L$ for
  $\vert\vec{q}\vert = 0.5$~GeV at densities $\varrho = 0$ (solid line),
  $\varrho = \varrho_0$ (long dashes) and $\varrho = 2\varrho_0$ (short
  dashes). The calculations have been performed using the parameters of 
  refs.~\cite{ChanfrayRappWambach, RappChanfrayWambach} 
  ($\Lambda = 1200$~MeV, $g'_{11} = 0.8$ and $g'_{12} = g'_{22} = 0.5$).}}
\end{figure}

In fig.~13 the longitudinal and transverse parts of the 
$\rho$ propagator are shown as a function of $M$ and for the
same densities and momenta as the selfenergies in fig.~12.
As expected, the width of the $\rho$ meson increases with density.
Defining the mass of the $\rho$ meson by the zero of the real part 
in the propagator (``pole mass''), we find an increase with density.
However, because of the strong broadening, at least at
$\varrho = 2\varrho_0$, the quasiparticle concept breaks down.
At this density one can also see that the spectral function, i.e. the
imaginary part of the propagator, develops two maxima, one above and one
below the vacuum $\rho$-meson peak. For $\vec{q}=0$ this was already
found in refs.~\cite{ChanfraySchuck,Herrmann}. At finite $\vec{q}$, however,
the transverse and the longitudinal propagator behave rather differently. 
%
%
\begin{figure}
\begin{center}
\epsfig{file=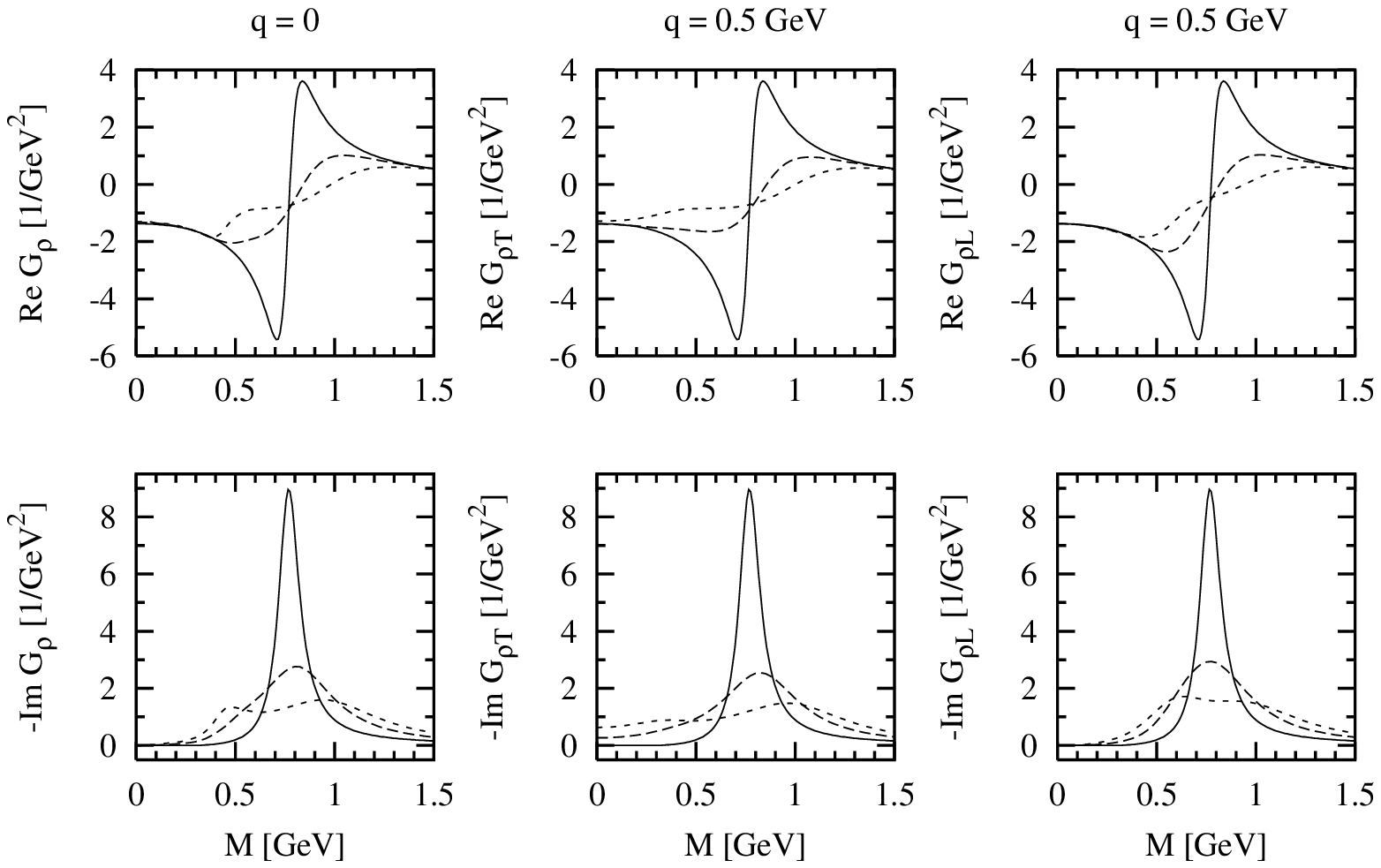,height=11cm,width=15.8cm}
\end{center}
\caption{\small {\it From left to right:
  $G_{\rho} = G_{\rho\,T} = G_{\rho\,L}$ for $\vert\vec{q}\vert = 0$,
  $G_{\rho\,T}$ and $G_{\rho\,L}$ for $\vert\vec{q}\vert = 0.5$~GeV at
  densities $\varrho = 0$ (solid line), $\varrho = \varrho_0$ (long dashes)
  and $\varrho = 2\varrho_0$ (short dashes). The calculations have been
  performed using the parameters of refs.~\cite{ChanfrayRappWambach,
  RappChanfrayWambach} ($\Lambda = 1200$~MeV, $g'_{11} = 0.8$ and
  $g'_{12} = g'_{22} = 0.5$).}}
\end{figure}

This can be seen even better in fig.~14 where the imaginary parts of
$G_{\rho\,T}$ and $G_{\rho\,L}$ at $\varrho = 2\varrho_0$ are plotted as
functions of $|\vec{q}|$ and $M$. The two branches we have seen in fig.~13
are clearly visible in both functions. Note that the lower branch of
$\mbox{Im}\,G_{\rho\,T}$ comes down to lower invariant masses with
increasing $|\vec{q}|$ and finally reaches $M=0$. This will become
important later on.   
%
%
\begin{figure}
\begin{center}
\epsfig{file=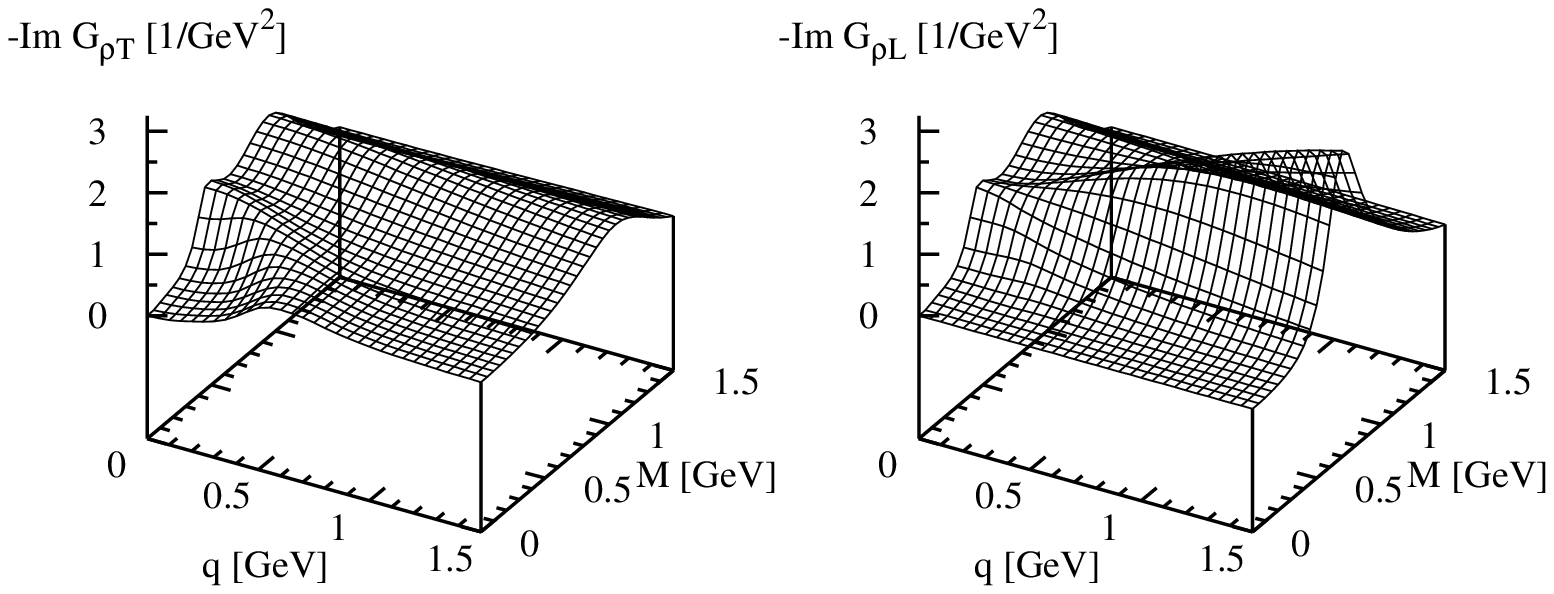,height=8.3cm,width=15.8cm}
\end{center}
\caption{\small {\it Imaginary part of the transverse and longitudinal $\rho$
  propagator $G_{\rho\,T}$ and $G_{\rho\,L}$ at $\varrho = 2\varrho_0$ as a
  function of the invariant mass $M$ and the momentum $\vert\vec{q}\vert$.
  The calculations have been performed using the parameters of
  refs.~\cite{ChanfrayRappWambach, RappChanfrayWambach}
  ($\Lambda = 1200$~MeV, $g'_{11} = 0.8$ and $g'_{12} = g'_{22} = 0.5$).}}
\end{figure}

For the interpretation we take the imaginary part of the propagator:
\beq
\mbox{Im} \;G_{\rho\,T,L}(q)={\mbox{Im}\;
\Sigma_{T,L}(q)\over{(q^2-(m_{\rho}^{(0)})^2-\mbox{Re}\,\Sigma_{T,L}(q))^2}
  + (\mbox{Im}\,\Sigma_{T,L}(q))^2}\ .
\label{ImGrho}
\eeq
Approximately, the maxima of $\mbox{Im} \;G_{\rho\,T,L}$ correspond to
maxima of $\mbox{Im}\;\Sigma_{T,L}$ or to minima of the denominator of
eq.~(\ref{ImGrho}). In vacuum only the latter are present.
With increasing density the vacuum peaks are pushed to higher
$q^2$ as a consequence of an increased real part of the selfenergy.
This explains the branches of $\mbox{Im} \;G_{\rho\,T}$ and 
$\mbox{Im} \;G_{\rho\,L}$ at higher $M$.
The lower branches correspond to $\mbox{Im}\;\Sigma_{T,L}$ and thus, 
according to eq.~(\ref{SigRhoMedijgen}), to the imaginary parts of the
functions $I_1$ to $I_7$. 

From the explicit expressions given in appendix~\ref{App3Niveau} 
(eqs.~(\ref{i13Niv}) - (\ref{i73Niv})) and neglecting the finite $\Delta$
width for the moment we find:
\beq
\mbox{Im}\,I_m(\vec{k},\vec{q},q_0) \;=\; \sum_{i,j=1}^3
  \varphi_{ij}^{(m)}(\vec{k},
\vec{q}) \; \delta( q_0^2 - (\omega_i(\vec{k}) + \omega_j(\vec{k}+\vec{q}))^2)
\qquad \rm{for \quad m = 1, ... , 4} \;,
\eeq 
\beq
\mbox{Im}\,I_6(\vec{k},\vec{q},q_0) \;=\; \sum_{i=1}^2 \sum_{j=1}^3
 \varphi_{ij}^{(6)}(\vec{k},\vec{q}) \; 
\delta( q_0^2 - (\Omega_i(\vec{k}) + \omega_j(\vec{k}+\vec{q}))^2)
\eeq 
and
\beq
\mbox{Im}\,I_5(\vec{k},\vec{q},q_0) \;=\; \mbox{Im}\,I_7(\vec{k},\vec{q},q_0)
  \;=\; 0 \;,
\eeq
where $\varphi_{ij}^{(m)}(\vec{k},\vec{q})$ are some real, well-behaved
functions. Including the $\Delta$ width basically leads to some 
``smoothing'' of the $\delta$-functions.

The structure of the expressions for $\mbox{Im}\,I_1$ to $\mbox{Im}\,I_4$
suggests the following physical interpretation: in matter the decay of the
$\rho$ meson into two pions is replaced by the decay into two longitudinal
quasiparticles with dispersion relations $\omega_i$ and $\omega_j$ 
(see eq.~(\ref{PiProp3Niveau})), i.e. $(Nh)_L$, $\pi$ or $(\Delta h)_L$ 
quasiparticles. The situation is slightly different for $I_6$ which
originates from a vertex correction to the $\rho\rho\pi\pi$-four point vertex
(see fig.~11). To this diagram also transverse $Nh$ and 
$\Delta h$ excitations contribute which have the dispersion relations 
$\Omega_1$ and $\Omega_2$. In fact, as explained in appendix~\ref{AppGen}, 
the decay into two longitudinal modes can be separated out and
$\mbox{Im}\,I_6$ describes the decay of the $\rho$ meson into one
longitudinal quasiparticle and one transverse $Nh$ or $\Delta h$ excitation.

Due to their small strength, $S_1$, the decay channels involving one or
two $(Nh)_L$ quasiparticles are relatively unimportant. The decay 
into two pion quasiparticles is suppressed by vertex corrections, as
already shown in refs.~\cite{ChanfraySchuck,KorpaPratt}. Therefore the most 
important
contributions to $\mbox{Im}\,I_1$ to $\mbox{Im}\,I_4$ come from the decay
$\rho\rightarrow\pi(\Delta h)_L$. Similarly $\mbox{Im}\,I_6$ is dominated by
the decay $\rho\rightarrow\pi(\Delta h)_T$. 

In order to identify these two modes in the spectral function we realize
that the pionic branch has most of its strength at low momenta (see fig.~3). 
Therefore a $\rho$ meson with momentum $\vec{q}$ should 
predominantly decay into a pion with momentum $\vec{k}\approx 0$ and a
$\Delta h$ pair with momentum close to $\vec{q}$. The corresponding
invariant masses 
\bea
M_{\rho\rightarrow\pi(\Delta h)_L}&\approx&
  \sqrt{\Big(m_{\pi}+\omega_3(\vec{q})\Big)^2-\vec{q}^{\,2}}\ ,
\nonumber \\
M_{\rho\rightarrow\pi(\Delta h)_T}&\approx&
  \sqrt{\Big(m_{\pi}+\Omega_2(\vec{q})\Big)^2-\vec{q}^{\,2}}\ , 
\label{Mrhopi}
\eea
are shown in fig.~15 for $\varrho = 2\varrho_0$. They are in good qualitative
agreement with the structures seen in the imaginary part of the 
$\rho$ propagator: We can identify the lower branch of $\mbox{Im}\,G_{\rho\,L}$
with the decay mode $\rho\rightarrow\pi(\Delta h)_L$ and the lower branch
of $\mbox{Im}\,G_{\rho\,T}$ with the decay mode 
$\rho\rightarrow\pi(\Delta h)_T$.
Therefore this structure of $\mbox{Im}\,G_{\rho\,T}$ is mainly a 
consequence of the vertex correction in fig.~11.
%
%
\begin{figure}
\begin{center}
\epsfig{file=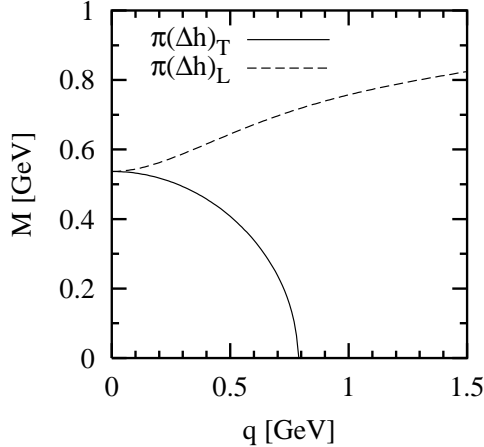,height=6.5cm,width=6.5cm}
\end{center}
\caption{\small {\it Invariant mass M of the $\rho$ meson decaying into
  a transverse (solid line) or longitudinal (dashed line) $\Delta h$ pair,
  resulting from the approximate formulae eq.~(\ref{Mrhopi}).}}
\end{figure}

All results presented so far have been calculated within the 3-level model,
i.e. the Fermi motion of the nucleons has been neglected.
In the final part of this section we want to test the quality of this
approximation.
Using spectral representations for the pion propagator $G_{\pi}$ and the 
selfenergy $\Pi^{\prime}$,
\bea
G_{\pi}(k_0,\vec{k})&=&-{1\over\pi}\int_0^{\infty} d\omega^2
  {\mbox{Im}\,G_{\pi}(\omega,\vec{k})\over
    {k_0^2-\omega^2+i\varepsilon}}\ ,
  \label{DispGpi}\\
\Pi^{\prime}(k_0,\vec{k})&=&-{1\over\pi}\int_0^{\infty} d\omega^2
  {\mbox{Im}\,\Pi^{\prime}(\omega,\vec{k})\over
    {k_0^2-\omega^2+i\varepsilon}}\ ,
  \label{DispPi}
\eea
the imaginary parts of the integrals $I_1$ to $I_4$ and $I_6$ reduce to
integrals from $0$ to $q^0$, e.g.
\beq
\mbox{Im}\,I_6(\vec{k},\vec{q},q_0) = -{2\over\pi}\int_0^{q_0} d\omega\;
  \mbox{Im}\,\Pi^{\prime}(\omega,\vec{k})
  \mbox{Im}\,G_{\pi}(q_0-\omega,\vec{k}+\vec{q})\ ,
  \label{I6disp}
\eeq
and similar expressions for $\mbox{Im}\,I_1$ to $\mbox{Im}\,I_4$,
while $\mbox{Im}\,I_5$ and $\mbox{Im}\,I_7$ vanish. From the structure of
these integrals one sees immediately that the imaginary part of the
$\rho$-meson selfenergy vanishes for $q_0\rightarrow 0$, as it should. 
As discussed above, because of the constant $\Delta$ width 
this was not exactly true for the results shown before. However, the
dispersion relations, eqs.~(\ref{DispGpi}) and (\ref{DispPi}), are also
slightly violated by the constant $\Delta$ width and the two effects cancel
to some extend giving the expressions like eq.~(\ref{I6disp}) the correct
behavior for $q_0\rightarrow 0$.

The expressions for $\mbox{Im}\,I_m$ and the remaining angular and
momentum integration in eq.~(\ref{SigRhoMedijgen}) can be evaluated
numerically. Some results for the imaginary part of the $\rho$-meson
selfenergy at
$\varrho=2\varrho_0$ are shown in fig.~16. Obviously the difference to the
results obtained within the three level model is small. This is somewhat
surprising since the 3-level model is known to be a quite poor
approximation to the pion selfenergy, at least in the domain of the
$N h$-excitations. However, after integrating over momenta, details of the
pion spectral function are ``washed out'' and only gross features are
important which are correctly reproduced by the 3-level model.
Of course the approximation becomes even better at lower densities where
the effect of Fermi motion is less important.
%
%
\begin{figure}
\begin{center}
\epsfig{file=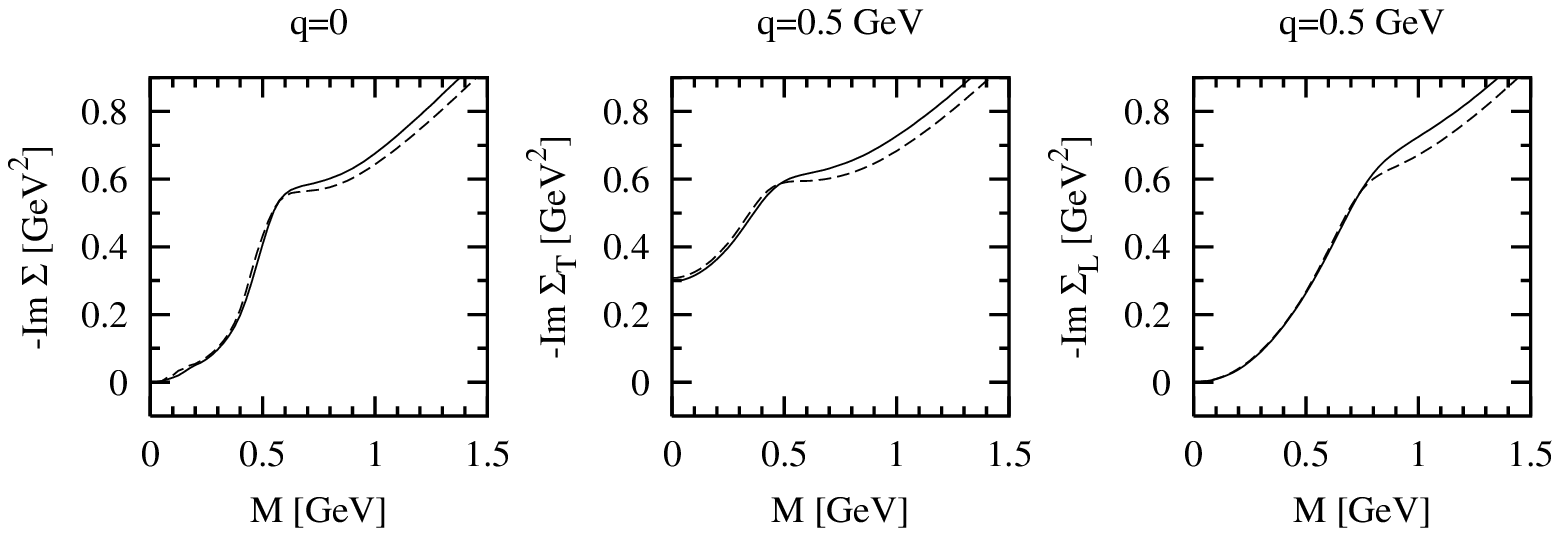,height=6cm,width=15.7cm}
\end{center}
\caption{\small {\it From left to right: Imaginary part of
  $\Sigma = \Sigma_T = \Sigma_L$ for $\vert\vec{q}\vert = 0$, $\Sigma_T$
  and $\Sigma_L$ for $\vert\vec{q}\vert = 0.5$~GeV, at density
  $\varrho = 2\varrho_0$.
  Solid line: full calculation, dashes: 3-level model.}}
\end{figure}
%
%
\section{Applications: Photoabsorption and Dilepton Production}

\label{KapApp}
\setcounter{equation}{0}
In this section we apply our model to the calculation of electromagnetic
processes, focusing on the effect of the 3-momentum dependence of the
pionic part of the $\rho$-meson selfenergy.   

We begin with the calculation of the total absorption cross section
of real photons on nucleons and nuclei \cite{RappUrbanBuballa}. In nuclear 
matter
the inclusive cross section per nucleon for a photon with momentum $\vec{q}$ 
and polarization vector $\epsilon_{\mu}(\vec{q},\lambda)$
is given by
\beq
{\sigma\over A} = {1\over{\varrho}}\,{e^2\over{2\vert\vec{q}\vert}}\sum_{f}
  \Big\vert\epsilon_{\mu}(\vec{q},\lambda)
  \bra{f}J^{\mu}(0)\ket{\Phi_0}\Big\vert^2
  (2\pi)^4\delta(p_f-q)\ .
\eeq
In this expression
$\ket{f}$ is the unobserved final hadronic state, $\ket{\Phi_0}$ specifies
the ground state of nuclear matter, and $J^{\mu}$ denotes the
electromagnetic current operator. 

This expression can be related to the current-current correlation function
\beq
J^{\mu\nu}(q)=-i\int d^4x\,e^{iq\cdot x}\bra{\Phi_0}T\Big(J^{\mu}(x)
  J^{\nu}(0)\Big)\ket{\Phi_0}\ .
\eeq
Inserting a complete set of energy and momentum eigenstates between the two
current operators and taking the imaginary part, we find
\beq
{\sigma\over A} = -{1\over{\varrho}}\,{e^2\over{\vert\vec{q}\vert}}\,
  \epsilon_{\mu}(\vec{q},\lambda)\epsilon_{\nu}(\vec{q},\lambda)
  \,\mbox{Im}\,J^{\mu\nu}(q)\ .
\eeq
In the VDM $J^{\mu}$ is a linear combination of the
$\rho$, $\omega$ and $\phi$ field operators. Neglecting the isoscalar
contributions, the current field identity reads
\beq
J^{\mu}(x)={(m_{\rho}^{(0)})^2\over g}\rho^{\mu}(x)\ ,
\eeq
i.e. $J^{\mu\nu}(q)$ is proportional to the $\rho$ propagator
$G_{\rho}^{\mu\nu}(q)$. Because of the transverse polarization of the photon
only $G_{\rho\,T}^{\mu\nu}(q)$ contributes, and we finally obtain
\bea
{\sigma\over A}=
  -{1\over{\varrho}}\,{e^2\over{\vert\vec{q}\vert}}\,
  {(m_{\rho}^{(0)})^4\over{g^2}}\,
  {\mbox{Im}\,\Sigma_T(q)\over{\vert(m_{\rho}^{(0)})^2+\Sigma_T(q)\vert^2}}\ .
\label{PhotoAbsWQ}
\eea
It is obvious that the 3-momentum dependence of the $\rho$-meson
selfenergy is necessary for the calculation of photoabsorption
cross sections within the VDM: Since $M=0$ for real
photons, $|\vec{q}|$ is the only variable the cross section can depend on.

For very small densities, $\varrho$, eq.~(\ref{PhotoAbsWQ}) describes the
photoabsorption cross section of a single nucleon. Since the $\rho$-meson
selfenergy in vacuum vanishes for $q^2=0$, the limit $\varrho\rightarrow 0$
is well defined and can be written as
\beq
\lim_{\varrho\rightarrow 0}{\sigma\over A} = -{e^2\over{\vert\vec{q}\vert}}\,
  {1\over{g^2}}\,\lim_{\varrho\rightarrow 0}
  {\mbox{Im}\Sigma_T(q)\over{\varrho}}\ .
\eeq
In this limit,
interferences between different contributions to $\Sigma_T$ vanish and
only contributions to $\Sigma_T$ linear in $\varrho$ survive, i.e. diagrams
containing exactly one $Nh$ or $\Delta h$ loop. Of course, describing
photoabsorption on a single nucleon, the creation of a hole is more
conveniently interpreted as the destruction of the initial nucleon.

Since the cross section is proportional to the imaginary part of $\Sigma_T$,
the processes which contribute correspond to cuts through the transverse
part of the $\rho$-meson selfenergy. Our model, as described in the previous
section, leads to the non-resonant background of the photoabsorption
cross section, e.g. 
$\gamma\rightarrow\pi+Nh$ or $\gamma\rightarrow\pi+\Delta h$. 
The threshold for such processes on a single nucleon at rest is 
$\vert\vec{q}\vert > 150$~MeV.

Our results for the non-resonant background of the photoabsorption cross 
section of the nucleon are shown in the left panel of fig.~17.
For comparison we also display the experimental data, averaged over protons
\cite{Armstrong-p} and neutrons \cite{Armstrong-n}. The short-dashed line
corresponds to a form factor cutoff $\Lambda=1200$~MeV as used in the Bonn
potential~\cite{BonnBibel,BonnPotential} and in
refs.~\cite{ChanfrayRappWambach,RappChanfrayWambach}. In this case, the
background alone exceeds the data for $q_0>800$~MeV, revealing the cutoff as
too large for the present purposes. The solid line was obtained with
$\Lambda=550$~MeV, which does not contradict the data.

Our results are in line with the well-known fact that for many applications
a much softer form factor is needed than the one used in the Bonn potential
(e.g. $\pi N$ scattering in the Chew-Low model~\cite{Layson}:
$\Lambda_{\pi NN} \approx 520$~MeV, $\pi^0$~photoproduction~\cite{MonizKoch}:
$\Lambda_{\pi N\Delta} = 300$~MeV, deep inelastic lepton
scattering~\cite{Frankfurt}: $\Lambda_{\pi NN(\Delta)} < 500$~MeV). In the
cloudy bag model for the nucleon~\cite{Thomas} the form factor has the form
$3 j_1(\vert\vec{k}\vert R)/(\vert\vec{k}\vert R)$, which for $R = 0.8$~fm
corresponds to a monopole with
$\Lambda_{\pi NN(\Delta)} = \sqrt{10}/R \approx 780$~MeV~\cite{BonnBibel}.
This value is similar to a result from lattice QCD~\cite{LiuDongDraper},
$\Lambda_{\pi NN} = (750\pm 140)$~MeV.
%
%
\begin{figure}
\begin{center}
\epsfig{file=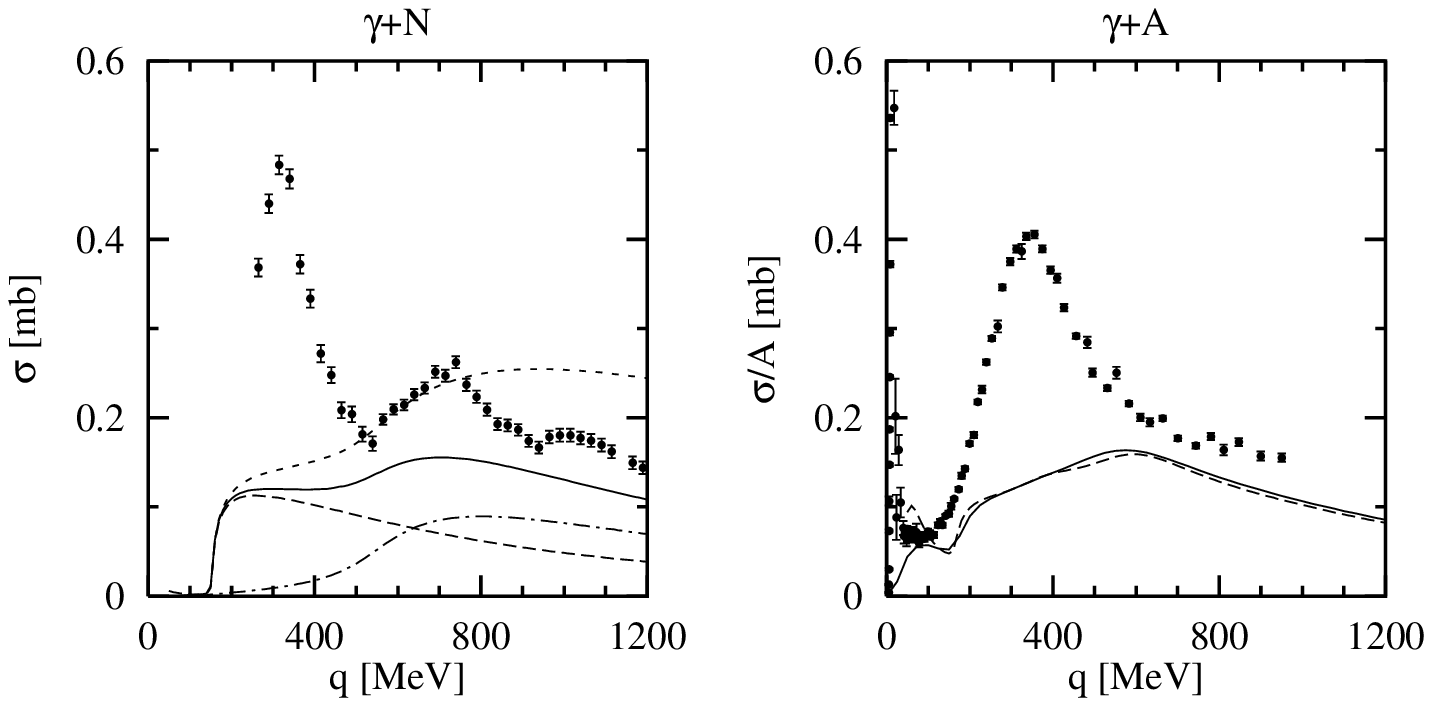,height=7.5cm,width=14.9cm}
\end{center}
\caption{\small {\it Left panel: Photoabsorption cross section of the
  nucleon. 
  The data are the averages of the total cross sections for protons  
  \cite{Armstrong-p} and neutrons \cite{Armstrong-n}. The theoretical
  calculations correspond to the non-resonant background only.
  They were obtained from eq.~(\ref{PhotoAbsWQ}) with 
  $\varrho=0.052\,\varrho_0$ and cutoff parameters $\Lambda = 1200$~MeV
  (short-dashed line) and $\Lambda = 550$~MeV (all other lines). 
  The background represented by the solid line is also separated into the 
  processes $\gamma N \rightarrow \pi N$ (long-dashed line) and 
  $\gamma N \rightarrow \pi\Delta$ (dashed-dotted line).
  Right panel: Photoabsorption cross section per nucleon of uranium.
  The data are averages of the data from
  refs.~\cite{Ahrens,Frommhold,Bianchi}, the background has been calculated
  using eq.~(\ref{PhotoAbsWQ}) with $\varrho=0.8\,\varrho_0$, 
  $\Lambda = 550$~MeV, $g'_{11} = 0.6$ and $g'_{12} = g'_{22} = 0.25$.
  The dashed line was obtained within the 3-level model. The solid line 
  corresponds to an improved calculation with the imaginary
  part of the $\rho$-meson selfenergy calculated with Fermi motion
  while for the real part the result of the 3-level model was taken.}}
\end{figure}

In fig.~17 we also show a decomposition of the background into the processes
$\gamma N \rightarrow \pi N$ (long-dashed line) and
$\gamma N \rightarrow \pi\Delta$ (dashed-dotted line). Since the
$\Delta$ resonance decays into $\pi N$ (implicitly contained in our model
through the constant width of the $\Delta$) the dashed-dotted line can
be identified with the two-pion background while the long-dashed line
corresponds to the single-pion background. The latter is somewhat larger
than the single-pion background found by Effenberger et al.
\cite{Effenberger} in an analysis of the $\gamma N \rightarrow \pi N$ data.
This might indicate that our formfactor should be reduced even
further
\footnote{The analysis of pion induced $\rho$-meson production 
($\pi N\rightarrow\rho N$), which was recently proposed by Friman 
\cite{Friman} as another test reaction, seems to give a similar 
result: In order to describe the data
\cite{Baldini} we would have to reduce the cutoff to about $300$~MeV.
However, in view of the rather difficult separation of the data from
competing $\pi N\rightarrow\pi\pi N$ processes, further careful studies
are required before one can draw firm conclusions.
See also ref.~\cite{RappMoriond}, where the photoabsorption cross section
is shown for $\Lambda=310$~MeV.}
.

For a realistic description of the photoabsorption 
resonance-hole excitations, like $\gamma\rightarrow \Delta h$ or
$\gamma\rightarrow N^*(1520)h$, have to be taken into account in addition
to the non-resonant background. 
The resonant contributions to the $\rho$-meson selfenergy have 
already been worked out in a momentum dependent way in 
ref.~\cite{RappChanfrayWambach} and can therefore easily be added to our 
model. This has been done in 
ref.~\cite{RappUrbanBuballa} and we refer to this publication for more 
details. 

In the right panel of fig.~17 we show our results for the photoabsorption
on nuclei together with the experimental data for uranium 
\cite{Ahrens,Frommhold,Bianchi}. The calculations have been done
using $\Lambda = 550$~MeV, $g'_{11} = 0.6$ and $g'_{12} = g'_{22} = 0.25$.
As an average density in finite nuclei we took 
$\varrho = 0.8 \varrho_0$. 
The dashed line was obtained within the 3-level model. The solid line 
corresponds to an improved calculation with the imaginary
part of the $\rho$-meson selfenergy calculated as outlined in the 
end of section \ref{KapRhoMed}, i.e. including Fermi motion. For the
real part we kept the result of the 3-level model. Obviously for energies
greater than $\sim 150$~MeV the differences between the two approximations
are small. At lower energies the differences in the selfenergy, although
still small in absolute units, are enhanced by the factor $1/|\vec q|$
in eq.~(\ref{PhotoAbsWQ}). Therefore the 3-level model should not be trusted 
in this regime.

In contrast to the photoabsorption on a single nucleon, in nuclear matter
there is no threshold at $|\vec q| \simeq 150$~MeV. For instance,
processes mediated by meson-exchange currents, like $\gamma \rightarrow
\pi\pi \rightarrow (Nh)(Nh)$ allow for the absorption of photons with
much lower energies. Of course our model cannot describe the giant
resonances at very low energies, but the pionic background contributes
considerably to the ``dip region'' at $|\vec q| \sim 150$~MeV.
 
Again, for a realistic description of the data resonant diagrams have
to be taken into account. More details about this can be found in 
ref.~\cite{RappUrbanBuballa}.

Finally we turn to the calculation of dilepton rates in hot nuclear matter. 
In the VDM the rate of $e^+e^-$ pairs with total 4-momentum $q$ produced 
per volume $d^3x$ is given by
\cite{ChanfraySchuck}:
\beq
{dN_{e^+e^-}\over{d^4xd^4q}}={\alpha^2\over{3\pi^3M^2}}\,
  {1\over{e^{q^0/T}-1}}\,{(m_{\rho}^{(0)})^4\over{g^2}}\,
  g^{\mu\nu}\,\mbox{Im}\,G_{\rho\,\mu\nu}^R(q;\,T)\ .
\label{DPR}
\eeq
Here $\alpha = e^2/(4\pi)$ is the fine structure constant, $M=\sqrt{q^2}$
the invariant mass of the dilepton pair, and $T$ the
temperature of the hadronic matter. 
$G_{\rho}^R(q;\,T)$ is the retarded $\rho$ propagator
at temperature $T$ and baryon density $\varrho$. 
Thus eq.~(\ref{DPR}) depends on temperature through the Bose factor
$1/(e^{q^0/T}-1)$ and through the retarded propagator. Since in
sec.~\ref{KapRhoMed} the $\rho$ propagator was calculated for $T=0$ only
we approximate 
\beq
\mbox{Im}\,G_{\rho}^R(q;\,T)\approx \mbox{Im}\,G_{\rho}(q;\,T=0)\ ,
\label{T0Naeherung}
\eeq
but keep the temperature dependence through the Bose factor.
(Note that at $T=0$ the imaginary parts of the retarded propagator
$G_{\rho}^R$ and the time-ordered one $G_{\rho}$ are equal.)

In fig.~18 the dilepton rates for $T=150$~MeV and $\varrho=\varrho_0$ are
shown. In the calculations we used the parameters obtained from the fit to
the photoabsorption spectra~\cite{RappUrbanBuballa}, 
i.e. $\Lambda = 550$~MeV, $g'_{11} = 0.6$ and 
$g'_{12} = g'_{22} = 0.25$. The solid line corresponds to our model
as described above. For $M \sim 0.2$ - $0.6$~GeV it is in fair
agreement with the `pion + one loop' result of Steele et al. \cite{Steele}.
However, whereas these authors find almost no medium effect at and above
the free $\rho$-meson mass, in our calculation the vacuum peak is strongly
reduced and we find enhanced rates at higher invariant masses. 
(For comparison see the dotted line which has been calculated using the 
$\rho$ propagator in vacuum.)
%
%
\begin{figure}
\begin{center}
\epsfig{file=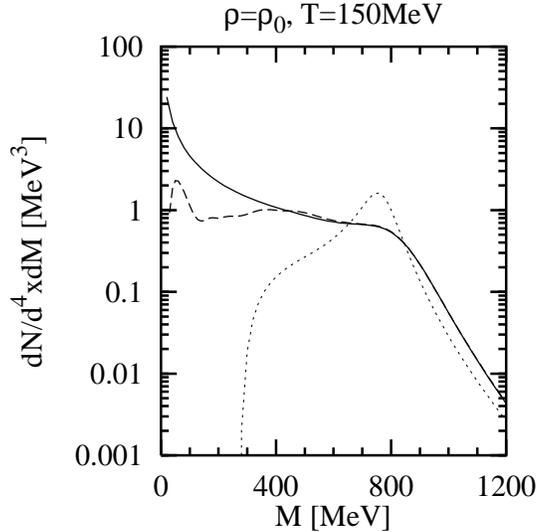,height=7.4cm,width=7.4cm}
\end{center}
\caption{\small {\it Dilepton rates for $\varrho=\varrho_0$ and $T=150$~MeV
  calculated from eq.~(\ref{DPR}) with different approximations to the
  retarded $\rho$ propagator $G_{\rho}^{R}(q^0,\vec{q};\,T)$:
  $G_{\rho}(q^0,\vec{q};\,T=0)$ (solid line), $G_{\rho}(M,\vec{q}=0;\,T=0)$
  (dashed line), and $G_{\rho}^\mathit{vac}(q^0,\vec{q};\,T=0)$ (dotted line).
  The parameters are
  $\Lambda = 550$~MeV, $g'_{11} = 0.6$ and $g'_{12} = g'_{22} = 0.25$.}}
\end{figure}

We also show the result of a calculation neglecting the
$\vec{q}$ dependence, i.e. using the approximation of
refs.~\cite{ChanfrayRappWambach,RappChanfrayWambach},
$\mbox{Im}\,G_{\rho}(q^0,\vec{q})\approx \mbox{Im}\,G_{\rho}(M,\vec{q}=0)$
(dashed line). Whereas the momentum dependence does not change very much for
larger invariant masses, for $M$ below $\sim 300$~MeV the results including
the $\vec{q}$ dependence of $G_{\rho}$ (solid line) are significantly enhanced
as compared to the calculation without $\vec{q}$ dependence (dashed line).
This can be explained as follows: As discussed in sec.~\ref{KapRhoMed}, the
$\rho$ meson strongly mixes with $\pi(\Delta h)_T$ states and the invariant
mass of these states decreases with increasing $\vec{q}$ and finally reaches
zero (see figs.~14 and 15). Because of the factor $1/M^2$ in eq.~(\ref{DPR})
this leads to a strong enhancement of the dilepton rate at small $M$. (Of
course, the divergence at $M=0$ is only an artifact of neglecting the electron
mass in the derivation of eq.~(\ref{DPR}).)   

Again, in a realistic calculation resonance contributions of the type 
$\rho \rightarrow B h$ have to be taken into account 
\cite{RappChanfrayWambach,FrimanPirner}. 
%
%
\section{Summary and Conclusions}

\label{KapSum}
\setcounter{equation}{0}
We have extended hadronic models for
$\rho$-meson propagation in cold nuclear matter involving
medium modifications of the pion propagator to include finite 
3-momenta. 
The starting point was the $\rho$ propagator in vacuum
with a bare propagator dressed by $\pi\pi$-loops. 
In matter, the pions are renormalized through particle-hole and
$\Delta$-hole excitations correlated by Migdal parameters.
We use a 3-momentum dependent monopole form factor to cut off
higher momenta at the $\pi NN$- and $\pi N\Delta$-vertices.
For numerical convenience, in most of our calculations
the dressed pion propagator was approximated by
the so-called ``3-level model'' which is obtained from the
exact propagator by neglecting the Fermi motion of the nucleons.
This approximation was found to be excellent for our purposes 
(see figs.~16 and 17). 

Because of the medium modifications of its pion cloud, the properties
of the $\rho$-meson become modified as well in matter. Guided by 
Ward-Takahashi identities, we have calculated vertex corrections to the
$\rho\pi\pi$- and $\rho\rho\pi\pi$-vertices in order to preserve
gauge invariance. For the resulting in-medium $\rho$ propagator
these vertex corrections turned out to be particularly important
to shift part of the strength down to the regime of low
invariant masses. The main effect could be attributed to the decay
of the $\rho$ into a pion and a spin-transverse $\Delta$-hole
excitation. For $|\vec q| = 0$ this branch was already found by
 Chanfray and Schuck \cite{ChanfraySchuck} and by Herrmann et al. 
\cite{Herrmann}. However, with increasing $|\vec q|$
it comes down to even lower invariant masses and eventually crosses
the $M=0$ line. 

As one application of the model we calculated the non-resonant
background for photoabsorption on nucleons and nuclei. 
In order to be consistent with the experimental data, the cutoff
parameter which enters the $\pi NN$- and $\pi N\Delta$-form factor
had to be reduced from $1200$~MeV as used in earlier calculations
\cite{ChanfrayRappWambach,RappChanfrayWambach} to $550$~MeV.
This example shows the importance of constraining the model parameters
from experimental data. As discussed in section \ref{KapApp}, 
there are indications from other observables, like more exclusive 
photoabsorption data  (see ref.~\cite{Effenberger}) or pion-induced 
$\rho$-meson production~\cite{Friman} that a further reduction of
the cutoff might be necessary. This is presently under investigation.
 
Finally we have studied the consequences of the 3-momentum dependence of 
the $\rho$-meson propagator for dilepton production rates in hot and
dense matter. Compared to a calculation where the $|\vec q|$-dependence
was neglected we found almost no difference for invariant masses
above $\sim 400$~MeV but a strong enhancement of the rates at lower
$M$. This is related to the $\pi (\Delta h)_T$-branch as discussed 
above.

In the present article we have concentrated on the pion-loop contribution
to the $\rho$-meson selfenergy. However, realistic calculations
require to account for contributions arising from
direct resonance formation $\rho \rightarrow B h$ 
\cite{RappChanfrayWambach,FrimanPirner,Peters}. In fact, the medium
modifications of the $\rho$ propagator associated with the pionic part of 
the selfenergy are substantially decreased by the use of much softer 
$\pi NN$ and $\pi N\Delta$  form factors. This increases the relative
importance of the resonances as compared 
to earlier versions of the model where a larger cutoff was used 
\cite{RappChanfrayWambach}. As long as interferences induced by explicit 
resonance decays are neglected, the resonance-hole diagrams 
can be evaluated separately and simply added to the pionic selfenergy 
contributions~\cite{RappUrbanBuballa}. Ultimately, one should take into
account the resonance decays into $\pi N$ and $\rho N$ in
a selfconsistent fashion~\cite{Peters}, which also implies a better
description of the $\Delta$ width. 
\vskip1cm
%
%
\centerline {\bf ACKNOWLEDGMENTS}
%
We are grateful for productive conversations with G.~Chanfray. 
One of us (RR) acknowledges support from the A.-v.-Humboldt-foundation 
as a Feodor-Lynen fellow and from the U.S. Department of Energy
under Grant No. DE-FG02-88ER40388.
\newpage
\begin{appendix}
%
\section{$\rho$ Meson Selfenergy at Finite Density}
%
\label{App}
%
\subsection{General Expressions for the Spatial Components}

\label{AppGen}
As described in sec.~\ref{KapRhoMed}, the spatial components of the
$\rho$-meson selfenergy are calculated by inserting the spatial components of
the vertex functions calculated in sec.~\ref{KapVertKorr} into
eq.~(\ref{SigRhoAllgemein}). If we split the vertex functions into bare
vertices and vertex corrections, as in eq.~(\ref{vertexcorrection}), the
selfenergy becomes a sum of seven terms:
\bea
\Sigma_{ij}(q)&=&
  {g^2\over{4}}\int {d^3k\over{(2\pi)^3}}\bigg(
  (2k+q)_i(2k+q)_j\,I_1 \nonumber\\
&&+4{(k_i(\Lambda^2-\vec{k}^2)-q_i\vec{k}^2)(2k+q)_j
  \over{\Lambda^2+(\vec{k}+\vec{q})^2}}\,I_2 \nonumber\\
&&+2{(k_i(\Lambda^2-\vec{k}^2)-q_i\vec{k}^2)
  (k_j(\Lambda^2-\vec{k}^2)-q_j\vec{k}^2)
  \over{(\Lambda^2+(\vec{k}+\vec{q})^2)^2}}\,\tilde{I}_3 \nonumber\\
&&+2{(k_i(\Lambda^2-\vec{k}^2)-q_i\vec{k}^2)
  (k_j(\Lambda^2-(\vec{k}+\vec{q})^2)+q_j\Lambda^2)
  \over{(\Lambda^2+(\vec{k}+\vec{q})^2)(\Lambda^2+\vec{k}^2)}}\,I_4
  \nonumber\\
&&+2\,I_5 \nonumber\\
&&+2{(\Lambda^2+\vec{k}^2)\delta_{ij}
  -(2k+q)_i(2k_j\Lambda^2-q_j\vec{k}^2)
  \over{(\Lambda^2+(\vec{k}+\vec{q})^2)^2}}\,I_6 \nonumber\\
&&-4\Big({\vec{k}^2\delta_{ij}\over{\Lambda^2+\vec{k}^2}}
  +{(2k+q)_i(k_j(\Lambda^2-\vec{k}^2)-q_j\vec{k}^2)
  \over{(\Lambda^2+(\vec{k}+\vec{q})^2)(\Lambda^2+\vec{k}^2)}}\Big)\,I_7
  \bigg)\nonumber\\
&&+(i\longleftrightarrow j)\ .
\label{SigRhoMedijalt}
\eea
with
\bea
I_1&=&{i\over\pi}\int dk^0\,G_{\pi}(k)G_{\pi}(k+q)\ ,\label{i1}\\
I_2&=&{i\over\pi}\int dk^0\,\Pi^{\prime}(k)G_{\pi}(k)G_{\pi}(k+q)\ ,
  \label{i2}\\
\tilde{I}_3&=&{i\over\pi}\int dk^0\,\Pi^{\prime\,2}(k)G_{\pi}(k)G_{\pi}(k+q)
\ ,\label{i3tilde}\\
I_4&=&{i\over\pi}\int dk^0\,\Pi^{\prime}(k)G_{\pi}(k)\Pi^{\prime}(k+q)
  G_{\pi}(k+q)\ ,\label{i4}\\
I_5&=&{i\over\pi}\int dk^0\,G_{\pi}(k)\ ,\label{i5}\\
I_6&=&{i\over\pi}\int dk^0\,\Pi^{\prime}(k)G_{\pi}(k+q)\ ,\label{i6}\\
I_7&=&{i\over\pi}\int dk^0\,\Pi^{\prime}(k)G_{\pi}(k)\ .\label{i7}
\eea
It is useful to rearrange these terms slightly~\cite{ChanfraySchuck}: 
in the integrals $I_2$, $\tilde{I}_3$, $I_4$ and $I_7$ the vertex corrections
$\Pi'$ are always coupled to a pion propagator of the same momentum.
Therefore, only the spin-longitudinal $Nh$ or $\Delta h$ excitations 
contained in $\Pi'$ can contribute to these integrals. This is different
for $I_6$. Here longitudinal and transverse $Nh$ and
$\Delta h$ excitations of the vertex correction can contribute, because they
are not coupled to a pion. However, we can separate
longitudinal and transverse $Nh$ and $\Delta h$ excitations in the $I_6$ term,
and combine the longitudinal part with the $\tilde{I}_3$ term. To that end, we
define the longitudinal spin-isospin response function,
\beq
\Pi_L^{\prime}(k)
=\Pi^{\prime}(k)+\Pi^{\prime}(k)\vec{k}^2G_{\pi}(k)\Pi^{\prime}(k)
=(k^2-m_{\pi}^2)\Pi^{\prime}(k)G_{\pi}(k)\ ,
\eeq
and an integral similar to $I_6$,
\bea
I_3={i\over\pi}\int dk^0\,\Pi^{\prime}_L(k)G_{\pi}(k+q)
=\vec{k}^2\tilde{I}_3+I_6\ .
\label{i3}
\eea
With this abbreviation we can write 
\bea
\Sigma_{ij}(q)&=&
  {g^2\over{4}}\int {d^3k\over{(2\pi)^3}}\bigg(
  (2k+q)_i(2k+q)_j\,I_1 \nonumber\\
&&+4{(k_i(\Lambda^2-\vec{k}^2)-q_i\vec{k}^2)(2k+q)_j
  \over{\Lambda^2+(\vec{k}+\vec{q})^2}}\,I_2 \nonumber\\
&&+2{(k_i(\Lambda^2-\vec{k}^2)-q_i\vec{k}^2)
  (k_j(\Lambda^2-\vec{k}^2)-q_j\vec{k}^2)
  \over{\vec{k}^2(\Lambda^2+(\vec{k}+\vec{q})^2)^2}}\,I_3 \nonumber\\
&&+2{(k_i(\Lambda^2-\vec{k}^2)-q_i\vec{k}^2)
  (k_j(\Lambda^2-(\vec{k}+\vec{q})^2)+q_j\Lambda^2)
  \over{(\Lambda^2+(\vec{k}+\vec{q})^2)(\Lambda^2+\vec{k}^2)}}\,I_4
  \nonumber\\
&&+2\,I_5 \nonumber\\
&&+2{(\Lambda^2+\vec{k}^2)^2
  \over{(\Lambda^2+(\vec{k}+\vec{q})^2)^2}}
  \Big(\delta_{ij}-{k_ik_j\over{\vec{k}^2}}\Big)\,I_6 \nonumber\\
&&-4\Big({\vec{k}^2\delta_{ij}\over{\Lambda^2+\vec{k}^2}}
  +{(2k+q)_i(k_j(\Lambda^2-\vec{k}^2)-q_j\vec{k}^2)
  \over{(\Lambda^2+(\vec{k}+\vec{q})^2)(\Lambda^2+\vec{k}^2)}}\Big)\,I_7
  \bigg)\nonumber\\
&&+(i\longleftrightarrow j)\ .
\label{SigRhoMedij}
\eea
%
%
\subsection{3-Level Model}

\label{App3Niveau}
Within the 3-level model the integrals $I_1$ to $I_7$ can be evaluated
analytically:
\bea
I_1
&=&\sum_{i=1}^{3}\sum_{j=1}^{3}{S_i(\vec{k})\over{\omega_i(\vec{k})}}\,
  {S_j(\vec{k}+\vec{q})\over{\omega_j(\vec{k}+\vec{q})}}\,
  {\omega_i(\vec{k})+\omega_j(\vec{k}+\vec{q})
  \over{q_0^2-(\omega_i(\vec{k})+\omega_j(\vec{k}+\vec{q}))^2}}\ ,
\label{i13Niv}\\
I_2
&=&\sum_{i=1}^{3}\sum_{j=1}^{3}{C_i(\vec{k})\over{\omega_i(\vec{k})}}\,
  {S_j(\vec{k}+\vec{q})\over{\omega_j(\vec{k}+\vec{q})}}\,
  {\omega_i(\vec{k})+\omega_j(\vec{k}+\vec{q})
  \over{q_0^2-(\omega_i(\vec{k})+\omega_j(\vec{k}+\vec{q}))^2}}\ ,
\label{i23Niv}\\
I_3
&=&\sum_{i=1}^{3}\sum_{j=1}^{3}
  {(\omega_i^2(\vec{k})-\omega_{\pi}^2(\vec{k}))
  C_i(\vec{k})\over{\omega_i(\vec{k})}}\,
  {S_j(\vec{k}+\vec{q})\over{\omega_j(\vec{k}+\vec{q})}}\,
  {\omega_i(\vec{k})+\omega_j(\vec{k}+\vec{q})
  \over{q_0^2-(\omega_i(\vec{k})+\omega_j(\vec{k}+\vec{q}))^2}}\ ,
\label{i33Niv}\\
I_4
&=&\sum_{i=1}^{3}\sum_{j=1}^{3}{C_i(\vec{k})\over{\omega_i(\vec{k})}}\,
  {C_j(\vec{k}+\vec{q})\over{\omega_j(\vec{k}+\vec{q})}}\,
  {\omega_i(\vec{k})+\omega_j(\vec{k}+\vec{q})
  \over{q_0^2-(\omega_i(\vec{k})+\omega_j(\vec{k}+\vec{q}))^2}}\ ,
\label{i43Niv}\\
I_5
&=&\sum_{i=1}^{3}{S_i(\vec{k})\over{\omega_i(\vec{k})}}\ ,
\label{i53Niv}\\
I_6
&=&\sum_{i=1}^{2}\sum_{j=1}^{3}{\alpha_i^{\prime}(\vec{k})
  \over{\Omega_i(\vec{k})}}\,
  {S_j(\vec{k}+\vec{q})\over{\omega_j(\vec{k}+\vec{q})}}\,
  {\Omega_i(\vec{k})+\omega_j(\vec{k}+\vec{q})
  \over{q_0^2-(\Omega_i(\vec{k})+\omega_j(\vec{k}+\vec{q}))^2}}\ ,
\label{i63Niv}\\
I_7
&=&\sum_{i=1}^{3}{C_i(\vec{k})\over{\omega_i(\vec{k})}}\ .
\label{i73Niv}
\eea
The abbreviations $C_i(\vec{k})$ are defined similarly to the strengths
$S_i(\vec{k})$ (see eq~(\ref{PiProp3Niveau})) by the equation
\beq
(\Pi^{\prime}G_{\pi})(k) = 
  {C_1(\vec{k})\over {k_0^2-\omega_1^2(\vec{k})}}
 +{C_2(\vec{k})\over {k_0^2-\omega_2^2(\vec{k})}}
 +{C_3(\vec{k})\over {k_0^2-\omega_3^2(\vec{k})}}\ .
\eeq

\end{appendix}
%
%

%
\end{document}